\documentclass[pdflatex,sn-nature]{sn-jnl}

\usepackage{graphicx}
\usepackage{amsmath,amssymb,bm}
\usepackage{booktabs}
\usepackage{array}
\usepackage{url}
\graphicspath{{figs/}}
\newcommand{\ci}[2]{[#1, #2]}
\newcommand{\dr}{\Delta r}
\unnumbered
\newif\ifappendSI
\appendSItrue%
\newif\ifsubmission
\submissionfalse
\ifsubmission
  \appendSIfalse
  \usepackage[nolists,nomarkers,figuresonly]{endfloat}
  \renewcommand{\includegraphics}[2][]{\centerline{\small[Figure file uploaded separately]}}
\fi

\begin{document}

\title[Solitary movement style predicts collective foraging]{Solitary movement style predicts collective foraging while whom to follow emerges only in company}

\author*[1]{\fnm{Hiroshi} \sur{Higashi}}\email{hgshrs@kansai-u.ac.jp}
\affil*[1]{\orgdiv{Faculty of Engineering Science}, \orgname{Kansai University}, \orgaddress{\city{Suita}, \state{Osaka}, \country{Japan}}}

\abstract{How far can behaviour inside a group, and the group's performance, be predicted from members measured while alone?
We tested this on three public datasets of human collective foraging (621, 128 and 1,133 participants).
A low-dimensional individual representation estimated from private-phase movement was mapped onto behaviour inside the group through a mapping learned on other participants.
Private information processing was stable between the two halves of a session (split-half reliability, the correlation between the halves, 0.60--0.83) and predicted behaviour inside the group with correlations of 0.30--0.66.
Selective social learning, following those who have succeeded or stopped, showed stable individual differences inside the group (reliability 0.45--0.59) yet was not predicted at all from the same person's private behaviour.
Group performance was predicted with a correlation of 0.55 in all three datasets by simulations never fitted to group outcomes or by the mean of individually measured performance, a prediction carried by mobility and individual performance rather than by social weights.
We found no credible evidence for an effect of how members were combined (equivalence bound 0.06).
Whom to follow thus emerges only in company, which delimits what solitary measurements can predict and compose.}


\maketitle

\section{Introduction}
Individual differences in behaviour are consistent across time and situations to some degree, and this consistency has been studied as behavioural syndromes or personality\cite{sih2004,reale2007,bell2009}.
The consistency of individual differences is, however, context dependent, and individuals also differ in the plasticity with which their behaviour changes between situations\cite{dingemanse2010,sih2015}.
Social behaviour is the typical case: the presence of others and the availability of information about them change both how an individual behaves and how individual differences are expressed\cite{webster2011,harcourt2009}.
Human use of social information shows consistent individual differences\cite{molleman2014,toelch2014}, yet it also changes flexibly with the situation\cite{witt2024}, and in animals personality predicts the role taken in social foraging\cite{kurvers2010}.
When research on collective intelligence asks whom to assemble and how to combine them\cite{couzin2009,krause2010,woolley2010,riedl2021,hong2004,page2007,bang2017,jolles2017,jolles2020,farine2017,barkoczi2016,galesic2023}, it implicitly assumes that the members' traits can be measured before they enter the group.
Yet which traits measured in advance, and alone, are preserved inside the group, and which emerge only in a social context, has not been tested under matched conditions.
The distinction is also practical.
Only traits that are expressed alone can be used to predict and compose groups from solitary measurements.
This study takes ``what can be measured in advance'' as its main question and, on that basis, tests whether the mean or the combination of the measurable traits determines group performance.

As the setting for this question we use three public datasets of human collective foraging.
Chirkov et al.\cite{chirkov2026} had 621 participants search for and track a moving resource in an immersive three-dimensional environment, and showed that groups outperform solitary individuals only when the others' payoff information is visible, and that the mechanism is payoff selectivity, following only those others who are doing better than oneself.
Payoff selectivity is a form of the ``learn from successful others'' strategy in the social-learning-strategy literature\cite{laland2004,rendell2010,rendell2011,kendal2018}; payoff-biased social learning is adaptive yet underused by humans\cite{mcelreath2008,mesoudi2011}, and how it is used has been shown to separate successful from unsuccessful groups in both experiments and models\cite{toyokawa2019,garg2022,deffner2024,mezey2024}.
When the others' payoffs are not visible, who knows has to be inferred from the others' behaviour\cite{heyes2016}, which is exactly the situation of the task of Hawkins et al.
In line with the framework that links cognitive computational models of individuals to collective behaviour\cite{tump2024}, their computational model estimated weights for each individual, but the explanation of group performance used simulations of homogeneous agents, and ``whether individual differences and their combination drive collective tracking'' was left as an explicit open question.
Recordings of the same kind are public for collective foraging in Minecraft\cite{wu2025} and for a two-dimensional collective sensing task in which the others' payoffs are not visible\cite{hawkins2023}, a human realisation of emergent sensing by animal groups\cite{berdahl2013}.
The three complement each other not only in the form of the task but in the presence of a truly solitary phase (Wu et al.\ and the waiting room of Hawkins et al.\ only) and in the visibility of the others' payoffs (some conditions of Chirkov et al.\ only).
In particular, the Chirkov et al.\ data alone cannot tell whether selective social learning ``cannot be measured'' or ``does not appear alone'', whereas Wu et al.\ and Hawkins et al., which contain solitary phases, can.

\begin{figure}[tbp]
\centering
\includegraphics[width=0.85\textwidth]{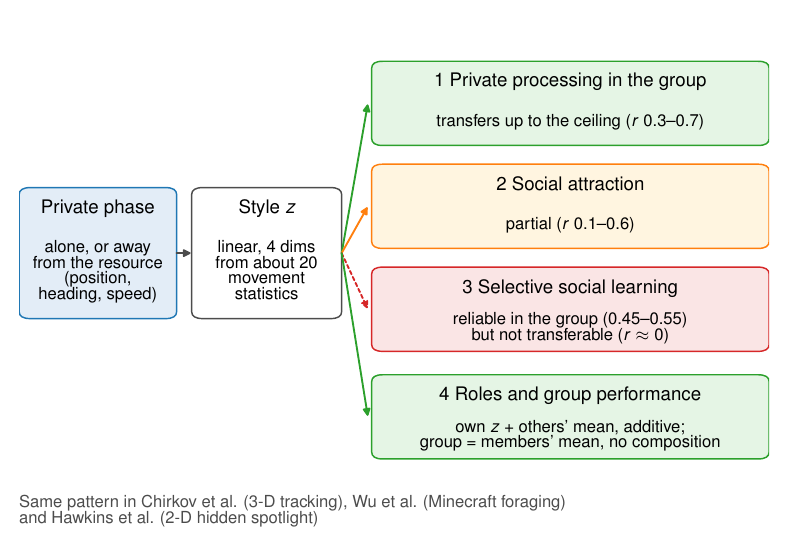}
\caption{\textbf{Individual style measured in private and the four directions in which its transfer was tested.}
A low-dimensional individual representation $z$ is estimated from behaviour recorded while a participant is alone (Wu et al.) or away from the resource (Chirkov et al., Hawkins et al.) and used to predict (1) private information processing inside the group, (2) social attraction, (3) selective (payoff- or state-dependent) social learning, and (4) roles in the attention network and group performance (group performance is predicted from individually fitted policies in Chirkov et al.\ and from the mean of individually measured performance in Wu et al.\ and Hawkins et al.).
Arrow colour summarises the result across the three datasets.
Green: transfers (private processing up to the split-half ceiling; roles follow an additive own-plus-others structure; group performance is predicted by the members' mean).
Orange: partial transfer.
Red dashed: no transfer although the trait is reliable inside the group.}
\label{fig:concept}
\end{figure}

The form of the test is the individuality-transfer framework\cite{higashi2025,higashi2026}.
A low-dimensional individual representation $z$ is estimated from behaviour recorded in one context, a mapping from $z$ to behaviour in another context is learned on other individuals, and the mapping is applied to an unseen individual.
That individual differences in complex behavioural policies can be compressed into a latent space of a few dimensions, and that distance in that space predicts transfer performance, has been shown on large-scale behavioural records from a competitive game\cite{higashi2026}.
Because the question is ``does the correspondence learned on other people hold for a new person'' rather than ``can a quantity measured in private be taken as the quantity in the social situation'', a failure to transfer can be decomposed into the measurement reliability of the target (the ceiling) and the context dependence of the trait, and a behavioural model conditioned on $z$ becomes a generative model of the individual that can simulate groups that never existed.
We estimated $z$ from movement statistics of the private phase (Wu et al.: rounds played alone; Chirkov et al.\ and Hawkins et al.: time spent searching away from the resource) and tested its transfer in four directions (Fig~\ref{fig:concept}):
(1) private information processing inside the group, (2) attraction to others, (3) selective social learning dependent on payoff or state, and (4) roles in the attention network (who has whom in view at each moment) and group performance.
The main hypotheses and decision rules were fixed in an analysis specification before the analysis, and hypotheses that failed are reported alongside those that held (Methods).

The results split by trait class.
Private information processing transferred close to the measurement ceiling in all three datasets and attraction to others transferred partially, whereas selective social learning, despite stable individual differences inside the group, was not predicted from the private phase and emerged only in company.
Group performance was predicted with a correlation of 0.55 by the mean of the members' individually measured traits, this prediction was carried by the mean of mobility and individual performance, and we found no credible evidence for an effect of how the members were combined.

\section{Methods}

\subsection{Ethics}
This study is a secondary analysis of anonymised public data and collected no new data.
The ethical approval of each dataset is described in the original publications\cite{chirkov2026,wu2025,hawkins2023}.

\subsection{Datasets and the private/social split}
Table~\ref{tab:datasets} summarises the three datasets.
Each contains, for the same participants, both a private phase in which the influence of others is small and a social phase in which they interact with others.

\paragraph{Chirkov et al.\ 2026.}
From the Open Science Framework (OSF, ZRHWT) we obtained the trajectories of each participant (position and heading recorded 10 times per second, with, for each second, the attention network of who has whom in view, the number of others who see each participant and the number of others each participant sees, and the social features), the feature table aggregated per second, metadata, and the analysis code and the code of the agent-based model (ABM).
Participants performed the 15-min task alone (A, 144 participants) or in groups of five, and the groups were divided into three conditions by the visibility of the others' payoff information.
With no payoff information (NP) the others' positions and movements are visible but not their payoffs.
With voluntary sharing (VP) each participant can choose at any moment to show their own payoff to the others at a cost of 1 point per second.
With full payoff information (FP) everyone's current payoff (proximity to the resource centre) is displayed at no cost as a coloured marker above the avatar.
These four conditions were crossed with resource speed (slow: 40\,\% of the avatar speed; fast: 80\,\%), and the original study reported that groups outperform solitary individuals only under FP (VP intermediate).
The analysed sample is the same 621 participants as in the original study (fast A/NP/VP/FP = 73/72/86/74, slow = 71/80/79/86), of whom the 477 in group conditions belong to 103 groups (65 groups of five and 38 groups of four; a group of four is one in which one member was excluded by the quality criteria, and the original study also retained them).
We reproduced the main results of the original study (the order of performance across conditions, the distributions of the number of others who see a participant and the number of others a participant sees, and their negative correlation, the number of discovery events 829/261, and the qualitative patterns of the ABM) before the analysis.
Time spent tracking on the resource (radius 40 m) was the social phase and time spent searching after losing the resource was the private phase.
The 1-Hz features (absolute rotation, state, change in the detector, distance, mean position angle and movement angle of others in the field of view, and the quality of social information) followed the definitions of the original study, with the angle normalisers set to geometric constants (the field-of-view half-width of 45.75$^\circ$, that is half of the 91.5$^\circ$ field of view of the original code, which we write as about 90$^\circ$ in this paper, and 180$^\circ$).
The 53 participants with less than 10 s of search phase (all in slow conditions) were excluded from the search-to-tracking transfer ($n = 568$).
Participants from all conditions were used for the transfer from the private to the social phase, the three group conditions (477 participants, 103 groups) for roles and group performance, and VP only for the exploratory analysis of voluntary payoff sharing.
``Within cell'' standardisation refers to the cells of payoff condition $\times$ resource speed.
The folds were fixed as 10 folds of whole groups stratified by condition $\times$ speed.

\paragraph{Wu et al.\ 2025.}
From the raw logs of the public repository (position and head direction of each player recorded 20 times per second) we extracted movement, block destruction and observed reward splashes for each 120-s round, and used the field-of-view reconstruction in Unity (who has whom in view).
There are 128 participants in 32 sessions of four, each with 8 solo and 8 group rounds (half in random and half in smooth environments, counterbalanced in blocks of four rounds).
The solo rounds, played alone by the same participants, are the private phase and the group rounds of four are the social phase.
The folds were 8 folds of 4 sessions.

\paragraph{Hawkins et al.\ 2023.}
From the public repository (Experiment 3) and the separately published archive of the authors (Experiment 1) we obtained the game logs (position and heading of each player recorded 8 times per second).
Experiment 1 comprises 294 complete 5-min games (682 participants, 1--6 players, goals set by clicking, binary payoff display, four pre-generated fields) and Experiment 3 comprises 223 complete 6-min games (451 participants, 1--5 players, gradient field, keyboard control, payoff displayed as a percentage), together with the waiting room before each game (about 80 s of truly solitary movement).
Experiment 2, in which 28 participants each played alone together with artificial agents whose movements were scripted in advance (a design for manipulating the use of social information), contains neither real human groups nor group performance, so it cannot serve the question of this study (transfer to roles and performance inside human groups) and was not analysed.
Searching outside the hidden scoring region was the private phase and scoring inside the region was the social phase (in Experiment 3, a field value $\ge 0.5$ counted as inside).
Individual measurements used the first half (ticks $< 1200$), the targets were the second half, and all variables were residualised within experiment $\times$ group size $\times$ field.
The folds were 10 folds of whole games.

The targets in the social phase were always the second half of the session (Chirkov: the remaining 12 min; Wu: the last two rounds of each block; Hawkins: the second half), and the predictive models were trained on folds that did not contain the target group, session or game.

\begin{table}[tbp]
\centering
\caption{\textbf{The three public datasets analysed.}
Group counts are the groups used in the analyses (Chirkov: groups with four or five analysed members; Hawkins: games with at least two players, in parentheses at least three).}
\label{tab:datasets}
\footnotesize
\setlength{\tabcolsep}{4pt}
\begin{tabular}{p{2.0cm}p{3.7cm}p{2.8cm}p{3.4cm}}
\toprule
Dataset & Task & Size & Private phase / social phase \\
\midrule
Chirkov et al.\ 2026 \cite{chirkov2026} & Track a moving resource for 15 min in an immersive 3-D environment. Alone (A) or in groups of five (no payoff information NP / voluntary sharing VP / full payoff information FP) $\times$ resource speed (slow / fast). 10-Hz recording; the attention network (who sees whom) reconstructed each second from a 90$^\circ$ field of view & 621 participants (144 alone, 477 in groups); 103 groups (65 of five, 38 of four) & Search phase (more than 40 m from the resource) / tracking phase (on the resource) \\
Wu et al.\ 2025 \cite{wu2025} & 16 rounds of 120 s searching for reward blocks in Minecraft (8 solo, 8 in groups of four; random / smooth reward environments). 20-Hz recording; field of view reconstructed in Unity & 128 participants; 32 sessions (groups of four) & Solo rounds / group rounds \\
Hawkins et al.\ 2023 \cite{hawkins2023} & 1--6 players search a 2-D screen for an invisible circular scoring region that moves, in games of 5--6 min. Experiment 1: binary payoff display; Experiment 3: graded payoff field and waiting-room recording. 8-Hz recording & Experiment 1: 682 participants, 294 games; Experiment 3: 451 participants, 223 games; group level 281 games (169) & Search state (outside the region) and waiting room / scoring state (inside the region) \\
\bottomrule
\end{tabular}
\end{table}

\subsection{Individual parameters and trait classes}
The individual parameters of each dataset are of two kinds: the weights of the original study's computational model fitted per participant (teacher models) and summary statistics of behaviour.
We grouped them into four trait classes (private information processing, social attraction, selective social learning, and being attended to; Supplementary Table 1).

``Selective social learning'' generalises to the three datasets the property that Chirkov et al.\ called payoff selectivity.
Payoff selectivity is the tendency, when the others' payoff information is visible, to use the social information only of others who are doing better than oneself and not to follow the rest.
In the tracking model of Chirkov et al., the weights on social cues (the others' position angle and movement angle) depend on the quality of social information (whether the visible others are closer to the resource than oneself) through interaction terms, and the two participant-level random effects of these interaction weights (SocPos$\times$SocQual, SocMov$\times$SocQual) served as the index of individual differences in payoff selectivity.
In the ABM policy fitted to the same data, the slope $k$ and threshold $x_0$ of the sigmoid that turns the quality of social information into the probability of following others play this role (Eq.~\ref{eq:psoc}).
In Wu et al., the others' reward events are visible as splashes, so the weight on choosing blocks near a player just seen to be rewarded (SuccessfulProximity) and the weight on choosing blocks near a player seen not to be rewarded (UnsuccessfulProximity) express the same property.
In Hawkins et al.\ the others' payoffs are not visible, so one has to infer that another is scoring from the fact that they have stopped, and the measures are facing stopped others, the degree to which stopped others are faced in preference to moving others (the selectivity index), and the attraction $b_{\mathrm{stop}}$ to stopped others in the goal-direction model.
In this paper we call the measures of the three datasets together ``selective social learning'' and write ``payoff selectivity'' when we refer to the interaction weights of Chirkov et al.\ and the $k$, $x_0$ of the ABM.
``Social attraction'' is the tendency to move towards others regardless of their success or state (the main-effect weights on position and movement direction in Chirkov et al., looking at others and nearest-neighbour distance in Wu et al., approach, distance, facing and $b_{\mathrm{cen}}$ in Hawkins et al.).
``Being attended to'' is the degree to which a participant is seen by others (in-degree), defined from the attention network of who has whom in view each second (a visual sensory network\cite{strandburgpeshkin2013,bastien2020}), and is paired with the degree to which they see others (out-degree); in Hawkins et al., where the field of view is unrestricted, it was defined from who is facing whom.

\paragraph{Chirkov et al.}
The tracking model of the original study describes the magnitude of rotation in each second on the resource with a truncated exponential distribution whose linear predictor contains private cues (the resource detector, that is the change in distance to the resource, and the resource distance) and social cues (the mean position angle and mean movement angle of others in the field of view, and the interactions of the latter two with the quality of social information).
There are six participant-level random effects: the intercept is how much a participant rotates while tracking (motor activity), the detector weight is how strongly they correct their heading when losing the resource (private information processing), the weights on the others' position and movement direction are social attraction, and the interactions with quality are payoff selectivity.
Because the brms\cite{burkner2017} fit of the original study took more than a day per model in our environment, we estimated the same model (features, fixed-effect structure, priors, likelihood) by empirical Bayes on a GPU (penalised maximum likelihood with a Laplace marginal-likelihood correction and a diagonal random-effect covariance) and compared it with the brms fit on 20\,\% of participants (121 participants; 4 chains $\times$ 1,500 iterations, $\hat R \le 1.015$, no divergences).
The correlations of the random effects of the same participants were 0.998 and 0.997 for the intercept and detector and 0.91--0.94 for the social weights.
In the ABM policy parameters (below), $I$ is the baseline rotation while tracking, $S$ the sensitivity to the detector, $k$ and $x_0$ payoff selectivity, $r_{\max}$ the upper bound of random rotation while searching, and $\pi$ the probability of not rotating in a second (persistence, the intermittency of movement).

\paragraph{Wu et al.}
The block-choice model of the original study explains the choice of the next block to mine as a weighted sum of the features of candidate blocks.
Locality is the tendency to choose nearby blocks, BlockVisibility the tendency to choose blocks in view, and RewardPrediction the weight on the reward predicted by a Gaussian process\cite{rasmussen2006} from one's own reward experience (asocial learning); these constitute private information processing.
SuccessfulProximity and UnsuccessfulProximity are selective social learning as described above, and the adaptation terms are the changes in each weight with the time elapsed since one's last reward.
Per-participant maximum-a-posteriori estimates (batch L-BFGS) with the authors' population posterior as the prior were obtained for all group rounds and for each half, and their correlations with the authors' Stan\cite{carpenter2017} posterior means were 0.86--0.997.
As summary statistics of behaviour we used movement in the solo rounds (path length, speed, cells traversed and so on), the attention network of the group rounds (out-degree, in-degree), nearest-neighbour distance, and the number of times another player's reward splash was observed.

\paragraph{Hawkins et al.}
The private measures are the latency from entering the scoring region to stopping and the fraction of time stopped inside the region, which express the style of staying put once a hit is found.
The social measures are approach events to others (facing one other at high speed for at least 0.5 s), nearest-neighbour and centroid distance, facing others (a 45$^\circ$ cone), facing stopped others and its selectivity index, and the time faced by others.
The goal-direction model (Experiment 1: each goal click; Experiment 3: the signed rotation each second, explained by the angle to the others' centroid $b_{\mathrm{cen}}$, to the nearest stopped other $b_{\mathrm{stop}}$, to the nearest other $b_{\mathrm{nn}}$, and their interactions with being inside the region) was estimated per participant with empirical-Bayes shrinkage.

Split-half reliability is the Pearson $r$ between values estimated separately on the first and second half of the session (Wu: rounds 1--2 and 3--4 of each block) and served as the ceiling for transfer (no Spearman--Brown correction\cite{spearman1910}).
The reliability of individual parameters estimated from behavioural tasks is generally lower than that of self-report\cite{enkavi2019}, and the interpretation of computational-model parameters depends on the reliability of their estimates\cite{eckstein2021}, so transfer correlations are always reported next to their ceiling.

\subsection{Individual representation $z$}
All $z$ are linear, with the standardisation and projection estimated on the training participants of each fold only.
The search-phase $z$ of Chirkov et al.\ is a 4-dimensional partial least squares (PLS) projection of the search-phase summary statistics (mean, variance and quantiles of rotation, mean response to the others' position, search range, search time) and a one-hot encoding of condition.
The first-3-min $z$ is a 4-dimensional principal component analysis (PCA) of 17 statistics of the participant's own movement (mean, standard deviation (SD) and 90th percentile of rotation, fraction of non-rotation, fraction of large rotations, lag-1 autocorrelation, reversal rate; fraction of time not moving; speed statistics; resource-distance statistics; tracking fraction; number of state switches) and contains no features of others in view.
The $z$ of Wu et al.\ is a 4-dimensional PCA of 21 movement and foraging statistics of the solo rounds (path length, speed, cells traversed, blocks opened, distance between blocks, rotation, indices of area-restricted search\cite{dorfman2022}, number of rewards and so on; explained variance 33/19/11/10\,\%).
The $z$ of Hawkins et al.\ is a 4-dimensional PCA, within experiment, of 12 statistics of the first-half search state (fractions of speed states, path length, rotation, search range, click rate, distance and rotation; 7 statistics in Experiment 3).
The pre-specified hypothesis required that the $z$ of a deep encoder taking the 1-s search-phase series as input (a three-layer one-dimensional convolutional neural network (CNN) or a gated recurrent unit (GRU), mean pooling, variational bottleneck $M \in \{2, 4, 8\}$, condition fed to the decoder only) outperform the linear baseline.
In Chirkov et al.\ we tried six learning rates from $10^{-6}$ to $10^{-3}$, 500, 10,000 and 50,000 epochs, a GRU, and $M = 2, 8$, none of which beat a linear ridge on the summary statistics (see Results), so all subsequent $z$ are linear.

\subsection{Transfer prediction and cross-validation}
Rather than using $z$ itself as the prediction, a mapping from $z$ to the target (ridge regression, regularisation chosen by cross-validation within the training fold) was learned on participants of folds that did not contain the target group, session or game, and applied to the $z$ of unseen participants.
Performance is the out-of-fold Pearson $r$ over the participants of the held-out fold, and the ceiling is the split-half reliability of the target.
The difference $\dr$ from the baseline (Chirkov: search time only, first-3-min score; Wu: solo reward; Hawkins: first-half score) was given a 95\,\% confidence interval and a two-sided $p$ by a cluster bootstrap (3,000 resamples) over groups, sessions or games.
For the test of cross-individual transfer (pre-specified hypothesis (ii)), for all ordered pairs $(k, l)$ within a held-out fold, the standardised log squared error of predicting participant $k$'s target from participant $l$'s $z$ with the fold decoder was regressed on the standardised $z$ distance $\|z_k - z_l\|$ or on the Euclidean distance of the private-phase summary statistics (random intercept per participant $k$).
Because the linear $z$ is a linear function of the summary statistics, the two distances are strongly correlated and a joint regression with both is collinear, so we compared the single slopes and log-likelihoods.

\subsection{Variance decomposition of roles}
The targets (in-degree, out-degree and their analogues) were standardised and entered into a mixed model (maximum likelihood) with a random intercept per group (game, session) and a performance control (Chirkov: first-3-min score; Wu: environment and solo reward; Hawkins: residualisation), to which the participant's own $z$ (4), the mean $z$ of the other members (4) and the within-group SD of $z$ (4) were added in that order, each block being evaluated by a likelihood-ratio test.
The role prediction of Chirkov et al.\ predicts the roles over the remaining 12 min from the $z$ estimated from the participant's own movement in the first 3 min (without features of others), and the pre-specified control is the first-3-min score.

\subsection{Agent-based model of Chirkov et al.\ and per-participant policies}
The ABM of the original study (the authors' public code, mesa 2.2.4, unmodified) is a continuous-space model that mimics the experimental environment in 1-s steps.
A resource (radius 40 m) starts from the centre of a 400 $\times$ 400 m square arena and moves by a L\'evy walk\cite{zaburdaev2015} (a heading drawn uniformly at random and a number of steps in that heading drawn as $\min(\mathrm{Pareto}(1) + 1, 100)$).
The resource speed is 40\,\% (slow) or 80\,\% (fast) of the avatar speed of 7 m/s.
Agents start on the resource (10 m from the centre) facing the centre and advance 7 m per second, changing only their heading.
Agents and the resource that approach the edge of the arena are turned back towards the centre.
In each second inside the resource (distance $d \le 40$ m) an agent earns $40 - d$ points (40 points per second at the centre, falling towards 0 at the edge).
Each second, every agent observes the change in distance to the resource (the detector $\mathrm{det}_t$, normalised to $[-1, 1]$ by the maximum relative speed of avatar and resource), the mean position angle $\theta_{\mathrm{pos},t}$ of others within a 90$^\circ$ field of view, and the quality of social information $q_t$, and switches between a tracking and a searching state at a distance of 40 m.
When following the private rule in the tracking state, the agent reverses its direction of rotation with probability $p_u = 0.35$ if the detector has worsened since the previous second and with probability $p_s = 0.15$ otherwise (values of the original study).
Each agent sets the magnitude of its rotation every second by the rule
\begin{equation}
|\Delta\theta_t| =
\begin{cases}
|\theta_{\mathrm{pos},t}| + \varepsilon_t & \text{w.p. } p_{\mathrm{soc},t} \text{ (towards others in view)},\\
I + S\,\mathrm{det}_t + \varepsilon_t & \text{w.p. } 1 - p_{\mathrm{soc},t}, \text{ tracking},\\
U_t + \varepsilon_t,\; U_t \sim \mathrm{Uniform}(0, r_{\max}) & \text{w.p. } 1 - p_{\mathrm{soc},t}, \text{ searching},
\end{cases}
\label{eq:policy}
\end{equation}
\begin{equation}
p_{\mathrm{soc},t} =
\begin{cases}
\dfrac{1}{1 + \exp\{-k\,(q_t - x_0)\}} & \text{when others are in view},\\
0 & \text{otherwise},
\end{cases}
\label{eq:psoc}
\end{equation}
where $q_t$ is the quality of social information (tracking: the difference between the mean distance of visible others and one's own distance; searching: 1 if a visible other is on the resource and $-1$ otherwise) and $\varepsilon_t \sim \mathcal N(0, \sigma^2)$ is rotation noise.
In addition, in each second the agent does not rotate with probability $\pi$ (persistence).
The original study showed, with groups of homogeneous agents, that only agents with high selectivity $k$ outperform solitary individuals.

For each participant we fitted $(I, S, \log k, x_0, \log r_{\max}, \log \sigma, \mathrm{logit}\,\pi)$ by maximum a posteriori from the participant's own 1-Hz rotation data only.
This fit used the rotations of the participant's whole session (search and tracking phases) and did not use the private-phase $z$.
The selectivity $k$ and threshold $x_0$ appear only in seconds in which others are visible and therefore cannot be estimated from the private phase, and the 53 participants in slow conditions have almost no search phase (less than 10 s).
The likelihood was a per-second mixture of folded normal distributions (a convolution of a uniform and a normal while searching), with seconds of no rotation ($|\Delta\theta| < 1^\circ$/s) treated as a point mass (treating them as a density drives $\sigma \to 0$).
Because the folded likelihood is invariant under $(I, S) \to (-I, -S)$ we constrained $I \ge 0$.
The prior was a Gaussian centred on the pooled fit of the participant's condition $\times$ speed cell (estimated on the participants of the other folds only).
The conversion between the data detector scale and the ABM scale was $S_{\mathrm{abm}} = S_{\mathrm{data}}(1 + v)$ ($v$ the ratio of resource speed to avatar speed).
In the simulations $\sigma$ was fixed at the original study's 10$^\circ$ (with the fitted value of about 40$^\circ$ the ABM agents cannot track) and $\pi$ was the mean of the fitted values of the group members.
Each group was simulated with agents carrying the policies estimated in the fold that excluded that group (four agents for groups of four), on the same resource trajectory, for 100 seeds and 900 steps, and the seed means of score, fraction of time tracking, tracking efficiency (points per second of tracking), in-degree and out-degree, their between-individual correlation (in--out correlation), median distance from the resource centre and the gain in attention received by resource discoverers (discoverer attention gain) served as predictions.
Nothing was fitted to group performance.
Evaluation used the Pearson $r$ between prediction and observation standardised within condition $\times$ speed, confidence intervals for $\dr$ by a group bootstrap, and a sign-flip permutation test on the per-group squared errors.
In total 202,200 ABM runs (about 1.1 CPU hours) were required.

\subsection{Controls for composition effects}
A composition effect was defined as ``the change $\dr$ in out-of-sample $r$ when within-group heterogeneity is added to a model that already carries the members' mean''.
In the simulations of Chirkov et al.\ we used the following controls.
(1) The default parameters of the original study and (2) the within-cell mean parameters of the training fold are the pre-specified controls, but they give every group in a condition the same prediction, cannot rank groups within a cell, and yield $r \approx 0$ by construction.
(3) The mean of the performance of each member simulated alone (the nominal group) measures the contribution of interaction.
(4) Random re-composition replaces the members of a group at random with participants from other groups of the same condition (50 re-compositions $\times$ 10 seeds); it is the pre-specified test of composition but also changes the group's mean parameters.
(5) Mean-preserving re-composition selects, from 1,000 random re-composition candidates, the 5\,\% whose within-group means of the six policy parameters (including persistence) are closest to those of the actual group, so that the members' mean is the same as the actual group's and only the membership differs.
(6) Group-mean agents replace all members of a group with identical agents carrying the mean parameters of that group's members, removing within-group spread entirely.
The difference between the prediction simulated with the actual members and (6) is the direct estimate of the composition effect, and adopting it as the substantive test was recorded as a decision made after seeing the results (2026-10-20).
In the regression on observed group performance, the increase in $R^2$ when the within-group SDs of the five per-agent policy parameters ($I$, $S$, $k$, $x_0$, $r_{\max}$; persistence is excluded because it is set to the group mean) were added to the group-mean parameters was evaluated by a permutation test of the SD block.

\subsection{Neural policy and vectorised environment}
To check that the conclusions do not depend on the form of the policy, we also used a neural policy whose rules are not written by hand.
We re-implemented the original ABM in vectorised numpy and confirmed its agreement with the mesa version on a grid of homogeneous agents (tracking time $r = 0.997$, efficiency $r = 0.996$; Supplementary Fig.~3).
The neural policy is a shared multilayer perceptron (2 layers $\times$ 64 units) that takes the per-second features of the original study (tracking flag, detector and its change, distance, presence of social information, number of visible others, signed position angle of others, movement direction of others, quality of social information), the rotation state of the previous second, and a 4-dimensional participant embedding $\bm{e}$ (fitted from the participant's own rotations over the whole session, separate from the individual representation $z$ estimated from the private phase), and outputs the probability of no rotation, the magnitude of rotation as a truncated exponential, and the probability of each rotation direction.
Whereas the ABM policy describes each participant by six interpretable parameters, the neural policy learns the mapping from features to rotation itself and the participant's individuality is carried only by the embedding $\bm{e}$.
In each fold the network was trained on the rotations of the training participants only, and the $\bm{e}$ of the participants of the held-out group were fitted from their own data with the network frozen.
The individual $\bm{e}$ improved the negative log-likelihood of the held-out second-half data over the population policy by 0.078 nats/s (85\,\% of participants).
Each group was simulated for 100 seeds, with the same controls as for the ABM policy.

\subsection{Grid ABM of Wu et al.\ and the mobility term}
The simplified grid ABM of the original study (20 $\times$ 20 cells, one cell per step, 9 steps to mine, 600 steps) was driven by each participant's five non-adaptive choice weights (population weights plus the participant's deviation), and each session $\times$ environment was simulated on the four environment files actually played $\times$ 25 seeds.
Because the grid ABM moves every agent by the same amount at each step, we also built a variant with a mobility term, in which the probability $p_i$ that agent $i$ moves on a given step was set to the mean solo-round path length divided by the maximum over participants (0.28--1.00).
The variants are the participant's own weights with their own $p_i$, the group-mean weights and $p_i$, the population weights with the participant's own $p_i$, and each member alone.
None of them is fitted to the group.

\subsection{Search over social rules}
In the vectorised environment, seven rules that change a single element of the social part of the ABM decision rule (turn towards the nearest other; judge the quality of social information from the best-informed other; scan widely with probability 20\,\% while searching; align with the others' heading instead of their position; use only others on the resource as a source of information; scale rotation while tracking by 0.8; and the combination of the last two) and the baseline rule were simulated for 103 groups $\times$ 50 seeds, and the reproduction of the levels and between-group variation of the attention-network metrics was evaluated.

\subsection{Statistical prediction of group performance (Wu, Hawkins) and selectivity measured inside the group}
In Wu et al., the group-round reward (standardised within environment) was predicted by the mean of the members' solo rewards (zero-shot) and by a leave-one-session-out ridge on the members' mean $z$ and solo reward (plus the within-group SD).
In Hawkins et al., the second-half group score was predicted by the mean of the members' first-half scores (zero-shot, 281 games with at least two players) and by a leave-one-game-out ridge on the mean $z$ and first-half score (plus the SD; 169 games with at least three players).
As an in-sample test, the difference in $R^2$ between means only and means plus SDs was evaluated by a permutation test of the SD block.
To test the objection that ``the effective trait simply cannot be measured in the private phase'', the within-group SD of the choice weights fitted on the first two rounds of each block in Wu et al., and of the first-half selectivity measures in Hawkins et al., were added to the mean models above to predict second-half group performance.

\subsection{Pooled equivalence test}
The $\dr$ of the three datasets were recomputed per unit (Chirkov: group; Wu: group round; Hawkins: game), and a cluster bootstrap stratified within dataset (5,000 resamples) gave the point estimate of $\dr$, its 95\,\% and 90\,\% confidence intervals and the one-sided 95\,\% upper bound.
The equivalence bound of the two one-sided tests (TOST) was the maximum absolute value of the 90\,\% confidence interval\cite{lakens2017}.
The primary contrasts were one per dataset (Chirkov ABM policy, Wu statistical model, Hawkins statistical model); secondarily, all five contrasts were pooled.

\subsection{Pre-specified hypotheses and deviations}
This study was not publicly preregistered.
Before the analysis, an analysis specification stating the hypotheses, decision rules, controls, computational budget and prohibitions was fixed on 2026-09-24 (it is contained in the first commit, 0a7c67d0, of the public analysis repository), and the original text together with a dated decision log was registered unmodified on OSF (\url{https://doi.org/10.17605/OSF.IO/7GAE9}).
The decision rules of the specification were applied as written and the verdicts are reported; controls added after seeing the results (mean-preserving re-composition, group-mean agents) and analyses not in the specification (the replications in Wu et al.\ and Hawkins et al., the pooled equivalence test, the test of selectivity measured inside the group, the mobility term, and the search over social rules) are marked as such (Supplementary Table 5).
The analysis followed the specification in the order replication and audit of the original studies, transfer, roles, and group prediction.
The six pre-specified hypotheses were: (i) the $z$ of a deep encoder of the search phase predicts the tracking-phase random effects with $\dr \ge 0.10$, $p < 0.01$ against the condition-mean baseline and outperforms a linear ridge; (ii) distance between individuals in $z$ space explains transfer error; (iii) the first-3-min $z$ predicts later roles with $\dr \ge 0.05$, $p < 0.01$ against the initial-score baseline; (iv) prediction is best in the FP condition; (v) agents with individual parameters predict group performance with $\dr \ge 0.10$, $p < 0.01$ against the default-parameter and mean-parameter controls; and (vi) reproduction is more accurate than the random re-composition control.
The most important prohibitions, which were respected, were never to fit policies or agent parameters to group performance and never to include members of the evaluated group in training.

\subsection{Execution of the analysis and use of AI}
The implementation and execution of the analysis code were carried out on the author's server by an AI coding agent (Claude Code, Anthropic, running the Claude Fable 5.1 model) following the analysis specification written by the author, and all judgements on the reports of each stage and all entries in the decision log were made by the author.
The author verified every number in this paper with a script that regenerates them from the result files.
The same AI was used as an assistant in drafting the manuscript, and the author checked and revised the whole text.

\subsection{Statistical analyses and reproducibility}
All correlations are Pearson $r$, out-of-fold for transfer and computed on values standardised within cell for group prediction.
Confidence intervals are 95\,\% percentile intervals of a cluster bootstrap with groups, sessions or games as the resampling unit (3,000 resamples for transfer and roles, 2,000 for group prediction, 5,000 for pooling); all tests are two-sided with $\alpha = 0.05$.
The $p$ of $\dr$ comes from the bootstrap distribution and the $p$ of the re-composition controls from a permutation distribution\cite{nichols2002} of 2,000 resamples.
Blocks of the mixed models were evaluated by likelihood-ratio tests; because the variance decomposition of roles was run on 9 outcome variables over the three datasets (6 roles, 2 performance measures, 1 facing stopped others), individual $p$ values just below 0.05 are not taken as evidence of an effect.
Negative results (composition effects) are interpreted through the upper bound of the equivalence test and we do not state that ``there is no effect''.
For reference, a prospective experiment that splits groups at the median of their mean trait, assuming a correlation of 0.55 between the trait mean and group performance, would give a performance difference of $d \approx 0.98$ and 80\,\% power with 18 groups per arm (36 groups if the true correlation is 0.40), whereas detecting a composition effect of correlation 0.06 at the same power would require about 2,200 groups (about 350 even at 0.15).
Sample sizes are given with each analysis and the exclusion criteria for participants and groups are as stated above.
Software: Python 3.12 (numpy, scipy, scikit-learn, statsmodels, PyTorch 2.11), mesa 2.2.4 (for running the original ABM only), R 4.4.3 and brms 2.23 (for validating the empirical-Bayes teacher only).
All random seeds are fixed and every number can be recomputed from the public data and public code.

\section{Results}

\subsection{The individual differences that can be measured reliably are concentrated in private information processing}
We computed the split-half reliability of the individual parameters in each dataset and grouped them into the four trait classes (Fig~\ref{fig:reliability}; all parameters in Supplementary Table 2).
The parameters of private information processing are highly reliable in all three datasets.
In Chirkov et al.\ the rotation intercept had $r = 0.76$ and the sensitivity to the resource detector $r = 0.60$, in Wu et al.\ Locality had $r = 0.83$ and BlockVisibility $r = 0.76$, and in Experiment 1 of Hawkins et al.\ stop latency had $r = 0.69$ and the fraction stopped inside the region $r = 0.64$.
Social attraction was moderate: looking at others in Wu $r = 0.59$, approach events in Hawkins $r = 0.61$ and nearest-neighbour distance $r = 0.59$, and the weight on the others' position in Chirkov $r = 0.30$.
By contrast, the payoff selectivity of Chirkov (the interaction weights with the quality of social information) was at the measurement limit with $r = 0.09$--$0.12$.
On the other hand, SuccessfulProximity in Wu had $r = 0.45$ and facing stopped others in Hawkins had $r = 0.55$ (Experiment 1) and $r = 0.59$ (Experiment 3), so selective social learning can be measured with moderate reliability inside the group.
The reliability of being attended to was low, $r = 0.22$--$0.24$.
Among the ABM policy parameters of Chirkov, the reliabilities were $r = 0.94$ for $x_0$, 0.84 for persistence, 0.70 for $k$, and 0.58--0.62 for $I$ and $S$.
All transfer correlations below are to be read against these reliabilities as ceilings.

\begin{figure}[tbp]
\centering
\includegraphics[width=0.6\textwidth]{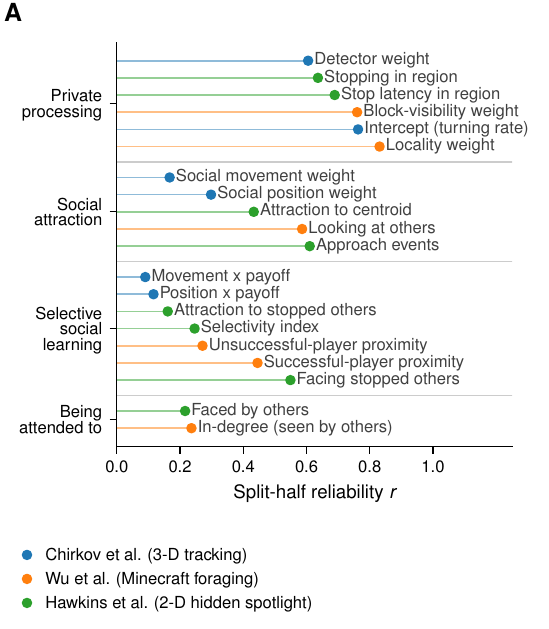}
\caption{\textbf{Split-half reliability of individual parameters by trait class.}
Split-half reliability of each individual parameter (Pearson $r$ between the first and second half of the session; Wu: rounds 1--2 vs 3--4 of each block), grouped by trait class.
Colour = dataset (legend below).
The figure shows representative parameters of each class; RewardPrediction in Wu (0.43) and Experiment 3 of Hawkins are not shown (all parameters in Supplementary Table 2).}
\label{fig:reliability}
\end{figure}

\subsection{The private-phase individual representation predicts private information processing close to the ceiling and does not predict selective social learning}
\begin{figure}[tbp]
\centering
\includegraphics[width=\textwidth]{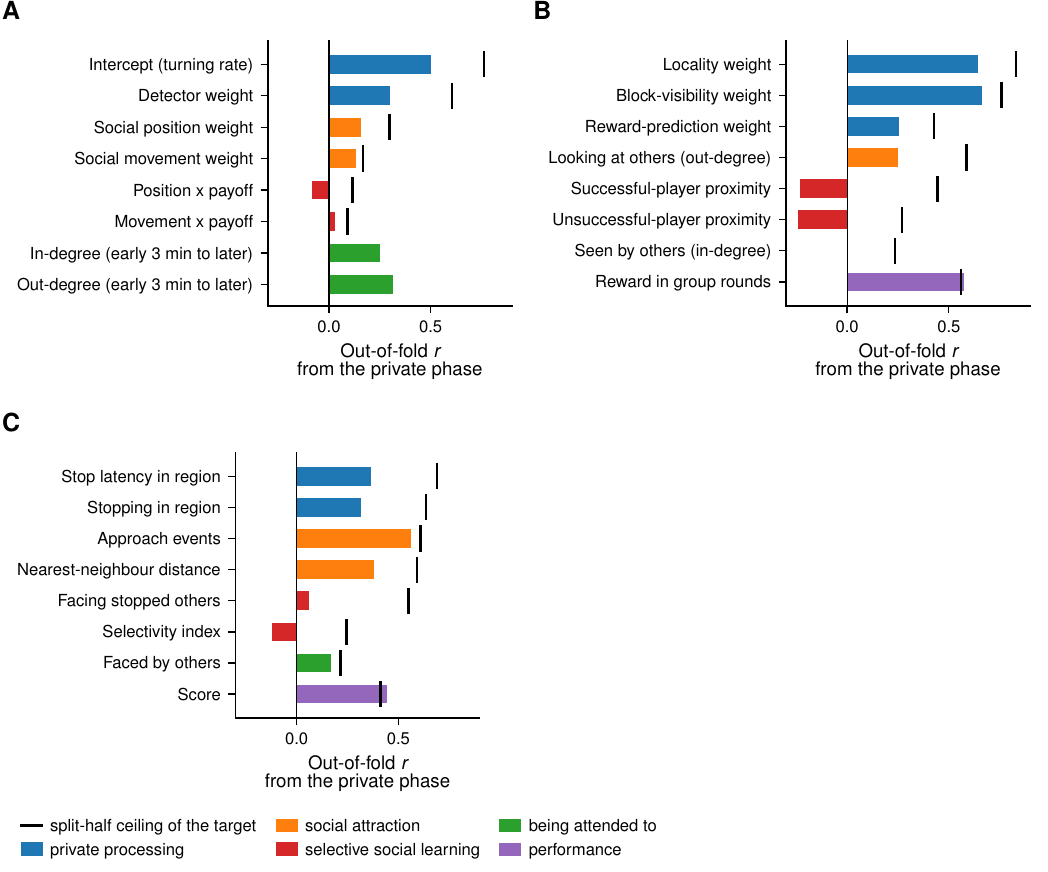}
\caption{\textbf{Transfer from the private phase to behaviour inside the group.}
Out-of-fold Pearson $r$ between the target measured in the social phase and its prediction from the private-phase $z$ (ridge regression; folds of whole groups, sessions or games).
Black ticks: split-half reliability of the target (ceiling).
Bar colour: trait class of the target (legend bottom right).
(A) Chirkov et al.: tracking-phase random effects predicted from the search-phase $z$ (PLS, 4 dimensions; $n = 568$), and later roles in the attention network predicted from the first-3-min movement $z$ (PCA, 4 dimensions; $n = 477$; no ceiling computed for roles).
(B) Wu et al.: group-round targets predicted from the solo-round $z$ (PCA of 21 movement and foraging statistics, 4 dimensions; 8 folds of 4 sessions; $n = 128$).
(C) Hawkins et al., Experiment 1: second-half targets predicted from the first-half search-state $z$ (PCA, 4 dimensions; 10 folds of games; $n = 682$; residualised within experiment $\times$ group size $\times$ field).}
\label{fig:transfer}
\end{figure}

A linear $z$ estimated from the movement statistics of the private phase was used to predict the individual parameters and behaviour of the social phase (Fig~\ref{fig:transfer}).
In Chirkov et al., the search-phase $z$ ($n = 568$) predicted the tracking-phase rotation intercept with $r = 0.50$ (ceiling 0.76) and the detector sensitivity with 0.30 (ceiling 0.60).
In Wu et al., the solo-round $z$ ($n = 128$) predicted the group-round Locality with $r = 0.64$ (ceiling 0.83), BlockVisibility with $r = 0.66$ (ceiling 0.76) and the weight on predicted reward with $r = 0.25$ (ceiling 0.43).
In Experiment 1 of Hawkins et al.\ ($n = 682$), the first-half search-state $z$ predicted the second-half stop latency with $r = 0.37$ (ceiling 0.69) and the fraction stopped inside the region with $r = 0.32$ (ceiling 0.64).
Performance itself also transferred (individual reward in the group rounds of Wu $r = 0.57$, ceiling 0.56; second-half score in Hawkins $r = 0.44$, ceiling 0.41).
No nonlinear $z$ was needed for this transfer.
The prediction of the tracking-phase random effects by the $z$ of a deep encoder reached only $r = 0.09$ on average over the six coefficients and did not beat a linear ridge on the summary statistics ($r = 0.17$) under any change of architecture, dimension, learning rate or number of epochs (pre-specified hypothesis (i) failed).

That $z$ functions as a coordinate system of individual differences was also confirmed by the cross-individual transfer error (pre-specified hypothesis (ii)).
In Chirkov et al., the error of predicting participant $k$'s tracking-phase parameters from participant $l$'s $z$ grew with the $z$ distance (38,498 pairs within held-out folds; slope 0.124, SE 0.003) and was explained better by it than by the distance of the summary statistics themselves (0.087; log-likelihood $-23{,}883$ vs $-24{,}505$).
The slopes were also positive in Wu et al.\ (1,920 pairs) and in Experiment 1 of Hawkins et al.\ (32,206 pairs) (0.267 and 0.208).
By target, the slopes were large for the targets that transfer (Locality 0.50 and BlockVisibility 0.57 in Wu; approach 0.52 and score 0.39 in Hawkins) and zero for the selective-social-learning targets that do not (SuccessfulProximity in Wu $-0.01$, $p = 0.65$; selectivity index in Hawkins $0.00$, $p = 0.91$).
For the traits that transfer, the property that distance in the latent space predicts transfer performance\cite{higashi2026} holds, in that people close in $z$ space can be predicted from one another, whereas for the traits that do not transfer, proximity in $z$ space carries no information.

Social attraction transferred partially.
In Experiment 1 of Hawkins, approach events were predicted with $r = 0.56$ (ceiling 0.61) and nearest-neighbour distance with $r = 0.38$ (ceiling 0.59); in Wu, looking at others with $r = 0.25$ (ceiling 0.59) and nearest-neighbour distance with $r = 0.34$ (ceiling 0.37); in Chirkov, the weight on the others' position with $r = 0.16$ (ceiling 0.30).

Selective social learning did not transfer.
In Wu, the prediction of SuccessfulProximity from the solo-round $z$ was $r = -0.23$ (ceiling 0.45) and that of UnsuccessfulProximity $-0.24$ (ceiling 0.27) (a negative out-of-fold $r$ is an artefact of averaging over folds when the prediction is close to constant and means the absence of signal).
In Experiment 1 of Hawkins, facing stopped others was predicted with $r = 0.06$ \ci{-0.02}{0.14} (ceiling 0.55) and the selectivity index with $r = -0.12$ (ceiling 0.25).
From the truly solitary 80 s of the waiting room before each game of Experiment 3, approach ($r = 0.20$) and score ($r = 0.26$) were predicted but facing stopped others ($r = -0.08$) and selectivity ($r = -0.16$) were not.
The attraction $b_{\mathrm{stop}}$ to stopped others in the goal-direction model was predicted from $z$ with $r = 0.33$ in Experiment 1, but its ceiling of $0.16$ is at the measurement limit, so this value cannot be interpreted and is not included in Fig~\ref{fig:transfer}.
The payoff selectivity of Chirkov cannot be evaluated for transfer because its ceiling is itself 0.09--0.12.
In Wu and Hawkins the same quantities have reliabilities of $r = 0.45$--$0.59$ inside the group, so selective social learning is a trait that ``does not appear alone'' rather than one that ``cannot be measured''.

\subsection{Roles in the attention network are determined additively by oneself and the others}
In Chirkov et al., the $z$ estimated from the participant's own movement in the first 3 min predicted the in-degree (being attended to) over the remaining 12 min with $r = 0.25$ and the out-degree (attending to others) with $r = 0.32$ ($n = 477$).
The control, the first-3-min score alone, gave $r = 0.14$ and $0.24$, and the increment of $z$ was $\dr = 0.11$ \ci{0.03}{0.20} ($p = 0.007$) for in-degree and 0.08 \ci{-0.03}{0.17} ($p = 0.14$) for out-degree (pre-specified hypothesis (iii) held for in-degree and failed for out-degree).
The pre-specified hypothesis (iv) that prediction would be best in the FP condition, where payoff information is visible, was not supported (in-degree $r$ 0.18/0.27/0.18 for NP/VP/FP).

\begin{figure}[tbp]
\centering
\includegraphics[width=0.55\textwidth]{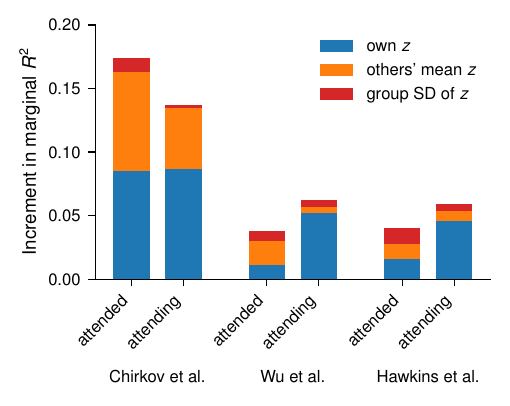}
\caption{\textbf{Variance decomposition of roles in the attention network (being attended to and attending to others).}
Increment in marginal $R^2$ of being attended to (in-degree) and attending to others (out-degree) when adding, in order, the participant's own $z$ (blue), the mean $z$ of the other members (orange) and the within-group SD of $z$ (red) to a mixed model with a group random intercept and a performance control, for the three datasets.
Likelihood-ratio $p$ values of the SD term are given in the text and in Table~\ref{tab:summary}.}
\label{fig:roles}
\end{figure}

Decomposing the variance of roles into the participant's own $z$, the mean $z$ of the other members and the within-group SD of $z$, in that order (Fig~\ref{fig:roles}), the own $z$ and the others' mean were effective in all three datasets and the within-group SD was not.
In Chirkov, the marginal $R^2$ of in-degree rose 0.08 → 0.16 → 0.24 → 0.25, with an SD increment of 0.01 (likelihood-ratio test $p = 0.22$), and 0.00 for out-degree ($p = 0.95$).
In Wu, attending to others was determined by the own $z$ ($R^2$ 0.002 → 0.054, $p < 0.001$) and being attended to by the others' mean (0.015 → 0.034, $p = 0.001$), and the SD contributed to neither ($p = 0.26$--$0.76$).
In Hawkins, too, the own $z$ ($p = 0.006$) and the others' mean ($p = 0.02$) were effective, and the SD increment of 0.012 for being attended to ($p = 0.048$) is not evidence of an effect once the comparison over 9 outcome variables is taken into account.
Among the outcome variables other than roles, only facing stopped others in Hawkins (selective social learning) retained an SD increment of 0.04 ($p = 0.003$) after correction, which suggests that the spread of movement styles within a group may relate to whom one follows, but a contribution to group performance was not supported by the composition tests below.
Roles are determined additively by ``one's own style'' and ``whom one is with (the others' mean)''.

\subsection{Group performance can be predicted from the mean of the members' individually measured traits}
\begin{figure}[tbp]
\centering
\includegraphics[width=\textwidth]{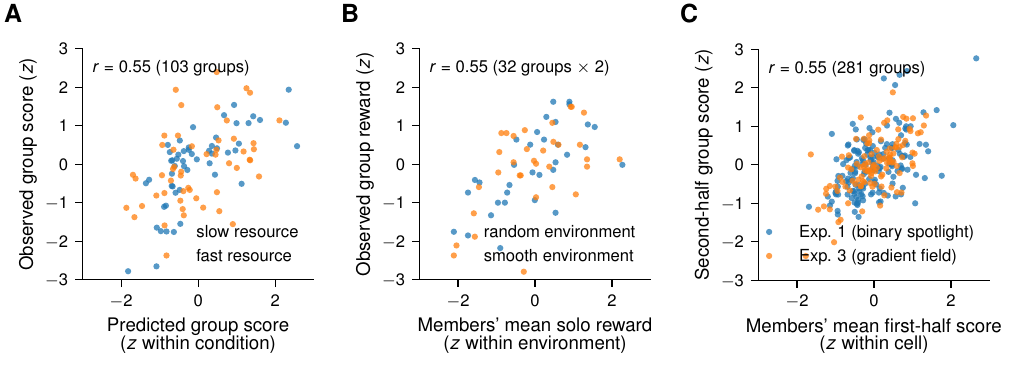}
\caption{\textbf{Prediction of group performance from the members' individually measured traits.}
Observed group performance against its prediction (both standardised within cell).
(A) Chirkov et al.: score of each group simulated with the members' individually fitted ABM policies (leave-one-group-out; mean of 100 seeds; 103 groups; blue slow, orange fast). $r = 0.55$ \ci{0.42}{0.67}.
(B) Wu et al.: group reward (mean of the four group rounds per session $\times$ environment, 64 cells) against the mean of the members' solo-round reward in the same environment. $r = 0.55$.
(C) Hawkins et al.: second-half group score against the mean of the members' first-half individual scores (281 games with at least two players; Experiment 1 blue, Experiment 3 orange). $r = 0.55$ \ci{0.44}{0.64}.
Nothing was fitted to group performance.}
\label{fig:group}
\end{figure}

In Chirkov et al., each group was simulated with agents carrying the ABM policy fitted to each participant's rotations over the whole session (Methods; Fig~\ref{fig:group}A).
The correlation between prediction and observation standardised within cell was $r = 0.55$ \ci{0.42}{0.67} for score, $r = 0.54$ \ci{0.40}{0.65} for tracking time, $r = 0.45$ \ci{0.26}{0.63} for tracking efficiency and $r = 0.41$ \ci{0.19}{0.57} for median distance (103 groups; the unstandardised correlation for score was $r = 0.75$; all metrics and all controls in Supplementary Fig.~1).
The pre-specified controls (the original study's default parameters, the within-cell mean parameters) gave $r \approx 0$ by construction ($-0.10$ and $-0.37$), so pre-specified hypothesis (v) held, but these controls cannot evaluate the effect of composition.
Against the mean of the performance of each member simulated alone, the simulation as a group was better by $\dr = +0.10$ \ci{0.04}{0.17} ($p = 0.001$; sign-flip permutation on the per-group squared errors $p = 0.016$) for score, so interaction contributed slightly to the prediction.
Under the pre-specified random re-composition the correlation of the fictitious groups was only $r = -0.02$ (95th percentile 0.14), but this is because re-composition also changes the group's mean parameters.
Under mean-preserving re-composition it recovered to $r = 0.43$ (95th percentile 0.52), and the difference from the prediction of the actual groups simulated with their own members ($r = 0.55$) was small (score $p = 0.012$, efficiency $p = 0.19$; pre-specified hypothesis (vi) held for random re-composition and failed for mean-preserving re-composition).
The neural policy, whose rules are not written by hand, gave the same conclusion, with score $r = 0.43$ \ci{0.24}{0.59}, tracking time $r = 0.38$ \ci{0.17}{0.55} and efficiency $r = 0.52$ \ci{0.32}{0.68} (Supplementary Fig.~2).

In Wu et al., the simplest predictor, the mean of the rewards the members obtained in their solo rounds, predicted the group-round reward in the same environment with $r = 0.39$ \ci{0.25}{0.52} per group round (256 rounds) and with $r = 0.55$ for the session $\times$ environment means (64 cells) (Fig~\ref{fig:group}B).
In Hawkins et al., the mean of the members' first-half scores predicted the second-half group score with $r = 0.55$ \ci{0.44}{0.64} (281 games; Fig~\ref{fig:group}C).
In all three datasets, the mean of the members' individually measured traits gave a correlation of 0.55 with the ranking of group performance, with nothing fitted to group performance.

\subsection{The prediction of group performance is carried by mobility and individual performance, and there is no evidence that individual differences in social weights contribute}
In Chirkov et al., the prediction that fell to $r = -0.02$ under random re-composition recovered to $r = 0.43$ under re-composition that matched the within-group means of the six policy parameters and reached $r = 0.54$ with group-mean agents, so the prediction is carried by the within-group means of the policy parameters.
Among them, re-composition that matched the means of the other five parameters but not that of persistence, the probability of not rotating in a second (reliability $r = 0.84$, second only to $x_0$), reached only $r = 0.38$ (95th percentile 0.48), so the contribution of the mean persistence can be identified as a difference in correlation of 0.05.
Which parameter means carry the rest cannot be separated in the present analysis.
In Wu et al., the individual reward in the group rounds could be predicted from the solo-round reward alone with $r = 0.61$ (ceiling 0.56), and $z$ ($r = 0.57$) did not exceed it.
At the group level, too, adding the mean $z$ to the mean solo reward ($r = 0.28$ \ci{0.05}{0.47}) changed nothing ($\dr = -0.01$ \ci{-0.09}{+0.08}), and the mean of the four private choice weights alone did not predict ($r = -0.30$).
The zero-shot simulation of the original grid ABM with each participant's choice weights was weak ($r = 0.20$ \ci{-0.03}{0.42}), and giving each agent a mobility drawn from its solo-round path length (reliability $r = 0.87$) raised it to $r = 0.44$ \ci{0.24}{0.64} ($\dr = +0.24$ \ci{0.04}{0.46}, $p = 0.02$; Supplementary Fig.~4, Supplementary Table 3).
Decisively, agents carrying the population choice weights and only their own mobility ($r = 0.41$ \ci{0.20}{0.60}) were indistinguishable from agents carrying their own choice weights and mobility ($r = 0.44$) ($\dr = +0.03$ \ci{-0.14}{+0.22}, $p = 0.71$).
In Hawkins et al., too, the increment from adding the mean $z$ to the mean first-half score ($r = 0.46$; 169 games) was only $\dr = +0.01$ \ci{-0.05}{+0.06}.
The component of group performance that can be predicted from prior measurement of individuals is the mean of ``how much one moves'' and ``how much one scores alone'', and individual differences in social weights did not change the prediction whether the participant's own values or the population values were used.

\subsection{There is no credible evidence of a composition effect, and its upper bound is an improvement in correlation of 0.06}
\begin{figure}[tbp]
\centering
\includegraphics[width=0.6\textwidth]{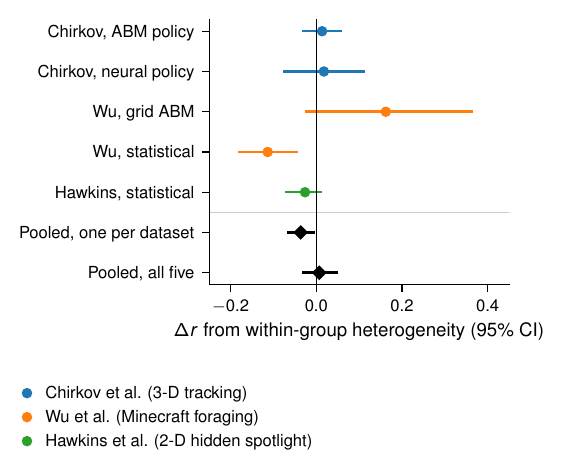}
\caption{\textbf{Change in out-of-sample prediction when within-group heterogeneity is added (composition effect).}
Change in out-of-sample $r$ ($\dr$, 95\,\% cluster-bootstrap confidence interval) when within-group heterogeneity is added to a model that already carries the members' mean traits.
Chirkov: individually fitted agents $-$ group-mean agents (ABM policy; neural policy; 103 groups).
Wu: the same contrast in the original grid ABM without the mobility term (64 session $\times$ environment cells), and member means + SDs $-$ means in a leave-one-session-out ridge (256 group rounds).
Hawkins: means + SDs $-$ means (169 games with at least three players).
Black diamonds: pooled estimates from a stratified cluster bootstrap over all units. One contrast per dataset (Chirkov ABM policy, Wu statistical, Hawkins statistical; $\dr = -0.04$ \ci{-0.07}{-0.01}) and all five contrasts ($+0.01$ \ci{-0.03}{+0.05}); one-sided 95\,\% upper bounds $-0.01$ and $+0.04$.}
\label{fig:composition}
\end{figure}

The composition effect (the $\dr$ when within-group heterogeneity is added to a model that already carries the members' mean) was estimated with matched controls in each dataset (Fig~\ref{fig:composition}, Table~\ref{tab:summary}).
In Chirkov et al., replacing all members of a group with group-mean agents still predicted score with $r = 0.54$, and the difference from the simulation with the actual members ($r = 0.55$) was $\dr = +0.01$ \ci{-0.03}{+0.06} ($p = 0.58$; sign-flip permutation $p = 0.62$).
Tracking time ($+0.01$), efficiency ($+0.02$ \ci{-0.02}{0.07}) and the neural policy ($+0.02$ \ci{-0.07}{+0.12}) behaved alike, and in the regression on observed performance the increase in $R^2$ from adding the within-group SDs was 0.03 ($p = 0.46$).
In Wu et al., adding the within-group SD to the members' means actually lowered the out-of-sample correlation from $r = 0.27$ to $0.16$ ($\dr = -0.11$ \ci{-0.18}{-0.04}; in-sample increase in $R^2$ 0.008, permutation $p = 0.77$; 256 group rounds), and in the grid ABM with mobility the difference between group-mean agents ($r = 0.47$) and the participants' own agents ($r = 0.44$) was $-0.02$ \ci{-0.16}{+0.12} ($p = 0.80$).
In the original grid ABM without mobility, the difference between the participants' own agents and group-mean agents leaned positive, $\dr = +0.16$ \ci{-0.02}{+0.36} ($p = 0.08$), but this reflects the weakness of the prediction itself without mobility ($r = 0.20$ vs $0.04$); Fig~\ref{fig:composition} and the pooling of all five contrasts below use this contrast without mobility.
In Hawkins et al., adding the SD to the means lowered $r$ from 0.47 to 0.44 ($\dr = -0.03$ \ci{-0.07}{+0.01}; in-sample increase in $R^2$ 0.03, permutation $p = 0.19$; 169 games).
Pooling the three datasets, one contrast per dataset (528 units, 304 clusters) gave $\dr = -0.04$ \ci{-0.07}{-0.01}, a one-sided 95\,\% upper bound of $-0.01$ and an equivalence bound of 0.06, and all five contrasts (695 units) gave $\dr = +0.01$ \ci{-0.03}{+0.05} with a one-sided upper bound of $+0.04$ and an equivalence bound of 0.04.
If a composition effect exists for the traits measured here, its size does not exceed the equivalence bound of the primary contrasts, 0.06 (0.04 for all five contrasts).

The objection that ``the effective trait, selective social learning, simply cannot be measured in the private phase'' was tested directly with selectivity measured inside the group (Supplementary Table 4).
In Wu, adding the within-group SD of SuccessfulProximity fitted on the first rounds did not change the prediction of the second-half group reward ($\dr = +0.01$ \ci{-0.07}{+0.05}; 128 group rounds), and in Hawkins the $\dr$ from adding the within-group SD of the first-half selectivity measures ranged from $-0.004$ to $-0.02$ (169 games; bootstrap $p = 0.63$--$0.91$, in-sample $\Delta R^2$ permutation $p = 0.17$--$0.81$).
For the spread of selectivity that can be measured reliably inside the group, too, we found no credible evidence that it improves group performance.

\begin{table}[tbp]
\centering
\caption{\textbf{Summary of the main results in the three datasets.}
Reliability is split-half reliability (two representative parameters). Transfer is the out-of-fold $r$ from the private-phase $z$ and its ceiling. Roles, SD: range of likelihood-ratio $p$ for adding the within-group SD in the variance decomposition. Group performance: the $r$ of Fig~\ref{fig:group}. $\dr$: the primary contrast of Fig~\ref{fig:composition} (95\,\% confidence interval).}
\label{tab:summary}
\footnotesize
\setlength{\tabcolsep}{4pt}
\begin{tabular}{p{3.9cm}p{2.65cm}p{2.65cm}p{2.65cm}}
\toprule
 & Chirkov et al. & Wu et al. & Hawkins et al. \\
\midrule
Reliability, private processing & 0.76 / 0.60 & 0.83 / 0.76 & 0.69 / 0.64 \\
Reliability, selective social learning & 0.09--0.12 & 0.45 / 0.27 & 0.55 / 0.25 \\
Transfer, private processing ($r$ / ceiling) & 0.50 / 0.76 & 0.64 / 0.83 & 0.37 / 0.69 \\
Transfer, selective social learning & not evaluable (ceiling $\le 0.12$) & $\approx 0$ (negative) & 0.06 \\
Roles: contribution of within-group SD ($p$) & 0.22--0.95 & 0.26--0.76 & 0.048--0.53 \\
Group performance, prediction $r$ & 0.55 (103 groups) & 0.55 (64 cells) & 0.55 (281 games) \\
Composition effect $\dr$ & $+0.01$ \ci{-0.03}{+0.06} & $-0.11$ \ci{-0.18}{-0.04} & $-0.03$ \ci{-0.07}{+0.01} \\
$\dr$ of the SD of selectivity measured inside the group & --- & $+0.01$ \ci{-0.07}{+0.05} & $-0.00$ \ci{-0.02}{+0.01} \\
Group-performance $r$ of the ABM with mobility & 0.55 (included from the start) & 0.44 (0.20 without mobility) & --- \\
\bottomrule
\end{tabular}
\end{table}

\subsection{What the policies do not capture: the structure of the attention network}
The policies above predict between-group differences in group performance but not between-group differences in the structure of the attention network (the bias in who looks at whom).
In Chirkov et al., the group mean of in-degree ($r = 0.00$ \ci{-0.27}{0.25}), the between-individual correlation of in-degree and out-degree ($r = 0.06$ \ci{-0.12}{0.25}) and the attention gain of resource discoverers ($r = -0.21$ \ci{-0.42}{0.03}) were predicted by neither the ABM policy nor the neural policy.
Seven rules that each change one element of the social part of the decision rule also left the between-group variation at $|r| \le 0.07$ (only the in--out correlation reached $r = 0.20$ under one rule), whereas the level of attention to discoverers (mean over groups) reached the observed 0.33 under rules that direct attention to others on the resource (0.31--0.41; Supplementary Fig.~5).
``In which groups attention becomes biased'' is determined by factors contained neither in the means of the individual parameters nor in uniform changes of the rule, and the effects of spatial configuration assumed by vision-based models of collective behaviour\cite{strandburgpeshkin2013,bastien2020} need to be considered in addition to the individual policies.

\section{Discussion}
The same structure emerged in three tasks.
The style with which a person processes information alone can be measured reliably from a few minutes of recording and transfers linearly to the social phase close to the measurement ceiling.
Whom to follow is expressed only inside the group; group performance is predicted by the mean of the members' individually measured traits, mainly their mobility and individual performance, and we found no credible evidence for an effect of how the members were combined.

\subsection{``Whom to follow'' is expressed only in a social context}
The central finding of this study is that the success of individuality transfer splits by trait class.
Selective social learning, following those who succeed or who have stopped, exists inside the group as individual differences with reliabilities of 0.45--0.59, yet cannot be predicted from the same person's private behaviour.
Because its reliability is at the measurement limit, the Chirkov et al.\ data alone could not tell ``cannot be measured'' from ``does not appear alone'', but Wu et al.\ and Hawkins et al.\ (including the waiting room), which contain truly solitary phases, showed the latter.
In the framework of behavioural syndromes\cite{sih2004,reale2007}, this is the boundary between individual differences consistent across contexts and those consistent only within a context, the latter corresponding to individual differences in the plasticity of behavioural reaction norms\cite{dingemanse2010,webster2011}.
Selective social learning is a trait expressed only once informed others exist; like leader--follower roles in pairs of fish\cite{harcourt2009} and leadership in moving groups\cite{couzin2005}, whom to follow is a product of interaction, consistent with the view of social learning strategies as switched according to context rather than fixed dispositions\cite{kendal2018,toyokawa2019}.
Groups therefore cannot be designed from solitary measurement with respect to whom to follow.
The payoff selectivity that Chirkov et al.\ identified as the key to collective intelligence\cite{chirkov2026} belongs precisely to this ``not measurable alone'' side.

\subsection{The predictable component of group performance lies in mobility and individual performance}
In all three datasets the mean of the members' individually measured traits predicted the ranking of group performance with a correlation of 0.55, and its content was simple.
In Chirkov et al., re-composition that matched the within-group means of the policy parameters recovered most of the prediction, with an identifiable contribution from the mean probability of not rotating; in Wu et al., solo-round mobility alone raised the prediction of the grid ABM from 0.20 to 0.44, and agents with the population choice weights and their own mobility were indistinguishable from agents with their own weights.
In Hawkins et al., too, adding the movement-style $z$ to the mean of the first-half individual scores did not change the prediction.
That is, zero-shot prediction from individuals to groups is almost entirely explained by the mechanism ``groups of people who move a lot and score well when measured individually perform well'', and we found no evidence that individual differences in social decision rules contribute to the prediction of group performance.
Mobility and individual performance are not independent: in Wu et al.\ the solo-round path length correlated with solo reward at $r = 0.24$ ($n = 128$); in Hawkins et al.\ the first-half path length in the search state correlated with the first-half score at $r = 0.35$ (Experiment 1, $n = 682$) and $0.27$ (Experiment 3, $n = 444$); and in Chirkov et al.\ persistence, the probability of not rotating in a second, correlated negatively with score ($r = -0.55$ in the solitary condition, $n = 144$; $-0.44$ in the group conditions, $n = 477$), where ``how much one moves'' is rotational activity because the avatars always advance at a constant speed.
People who move more may encounter the resource more often, but the pathway mobility → individual performance → group performance was not tested directly here.
In psychological terms, the individual differences that predict group performance in advance lie in general dimensions such as activity\cite{reale2007}, task engagement and the balance between exploration and exploitation\cite{monk2018}, not in individual differences in the use of social information.
This agrees with the finding that personality predicts roles in animal social foraging\cite{kurvers2010}, and adds to the finding of consistent individual differences in human social learning strategies\cite{molleman2014,toelch2014} that these differences can be measured only inside the group.

\subsection{Additivity and the ``diversity bonus'': a negative result as an upper bound}
The independence of the collective-intelligence factor $c$ from the members' mean ability\cite{woolley2010,riedl2021}, the possibility that diverse groups outperform groups of high ability\cite{hong2004,page2007} and the effects of individual heterogeneity on the movement and function of animal groups\cite{jolles2017,jolles2020,farine2017} are cited as grounds for the importance of how members are combined.
This study tested the hypothesis directly with matched controls and obtained no credible evidence of a composition effect.
In Chirkov et al.\ replacing all members with group-mean agents did not change the prediction accuracy for any policy type, and in Wu et al.\ and Hawkins et al.\ adding the within-group SD worsened the out-of-sample prediction.
The variance decomposition of roles showed the same additive structure: whom one is with changes one's role, but through the others' mean.
From the equivalence test\cite{lakens2017} pooling the 528 units of the three datasets (groups, group rounds and games; 304 clusters), a composition effect, if it exists, does not exceed the equivalence bound of 0.06.

This negative result is an upper-bound claim, ``any effect is $\dr \le 0.06$'', not a claim that ``there is no effect'', and the bound is restricted to the traits measured (movement style, response to the resource, simple social attraction, and selectivity measured inside the group) and to foraging and tracking tasks of 2--15 min.
Most reports of a diversity bonus concern cognitive problem-solving tasks in which diversity of knowledge and perspective is effective\cite{hong2004,woolley2010}, and group synergy depends on task complexity\cite{almaatouq2021}, so the same effect need not be expected in spatial foraging and tracking.
The objection that ``the effective trait simply was not measured'' does not hold, at least in Wu et al.\ and Hawkins et al., because the spread of selectivity measured inside the group did not improve group performance either.

\subsection{The role and limits of the individuality-transfer framework}
The individuality-transfer framework\cite{higashi2025} was effective here in the form of the question and of the test.
Measuring whether the mapping from $z$ to behaviour learned on other people applies to a new person, against the reliability of the target as a ceiling, separated failures to transfer into measurement limits and context dependence, and policies conditioned on per-individual parameters or embeddings made it possible to simulate groups that never existed, such as group-mean agents and re-compositions.
On the other hand, the deep encoder that the framework envisages did not beat linear regression on the Chirkov et al.\ data (568 participants; not tried on the other two datasets), and in the prediction of group performance $z$ added nothing to the mean of individually measured performance.
On these data the useful dimensionality of an individual representation is set by the reliability of the targets rather than by the expressive power of the encoder, in line with the general cautions on using computational-model parameters as indices of individual differences\cite{enkavi2019,eckstein2021}.

\subsection{Limitations}
This study is a secondary analysis of public data; its conclusions are limited to ``can be predicted'' and do not show ``can be manipulated''.
A prospective experiment composing groups from individually measured traits can test the effect of the mean with about 18 groups per arm (Methods), but lowering the upper bound of the composition effect would need about 2,000 groups.
All three tasks are 2--15 min foraging and tracking tasks in virtual space; whether the conclusions extend to long collaborations involving conversation or to foraging in the field\cite{schakowski2026} is unknown.
The selective-social-learning weights of Chirkov et al.\ are at the measurement limit, group prediction in the fast-resource condition is weak (neural-policy score $r \approx 0.12$), and the groups of four (38 groups) were simulated with four agents without the excluded fifth member.
The ABM under-predicts the absolute tracking time (0.6--0.7 vs 0.9 observed in slow FP) and uses the original study's rotation noise rather than the human fitted value, so evaluation was restricted to relative prediction within cell.
Only the solo rounds of Wu et al.\ are truly solitary measurements; the policies of Chirkov et al.\ were fitted to the whole session and the first-half scores of Hawkins et al.\ come from inside the group (and are confounded with practice effects), so only in Wu et al.\ was group performance predicted from individuals measured alone.

\subsection{Conclusions}
We have delimited the individual differences that prior measurement of individuals can capture.
Private information processing and mobility can be measured from short private-phase records, predict behaviour inside the group close to the measurement ceiling, and predict group performance through the mean of individually measured traits, whereas selective social learning, whom to follow, is expressed only inside the group.
Tests with matched controls on three datasets found no credible evidence for an effect of how members are combined, with an upper bound at the equivalence bound of 0.06.
In short foraging and tracking tasks, the component of group performance that can be predicted from prior measurement of individuals is thus additive and simple, whereas whom to follow and who looks at whom emerge only in company and must be measured there.

\backmatter

\bmhead{Supplementary information}
Supplementary Figs~1--5, Supplementary Tables~1--6 and Supplementary Note~1 are provided as a separate file.

\bmhead{Acknowledgements}
We thank Chirkov et al., Wu et al.\ and Hawkins et al.\ for making their data and code public.

\bmhead{Funding}
This work was supported by JSPS KAKENHI Grant Number JP24K15047.

\section*{Declarations}

\bmhead{Data availability}
All data used in this study are public.
Chirkov et al.\ 2026: OSF \url{https://doi.org/10.17605/OSF.IO/ZRHWT}.
Wu et al.\ 2025: \url{https://github.com/charleywu/minecraftforaging}.
Hawkins et al.\ 2023: \url{https://github.com/hawkrobe/emergent-sensing} (the raw ticks of Experiment 1 are in the authors' S3 bucket).

\bmhead{Code availability}
The analysis code, the pre-specified analysis plan, the decision log and all aggregate results are available at \url{https://github.com/hgshrs/collective-individuality-transfer} and archived on Zenodo (DOI 10.5281/zenodo.23582821).
The analysis specification and decision log are also registered on OSF (\url{https://doi.org/10.17605/OSF.IO/7GAE9}).
Participant-level raw data are not redistributed and are re-created from the public datasets of Chirkov et al.\ (OSF ZRHWT), Wu et al.\ (github.com/charleywu/minecraftforaging) and Hawkins et al.\ (github.com/hawkrobe/emergent-sensing) with the scripts provided.
The per-participant estimated parameters, the raw simulation outputs and the submission figure files are archived separately on Zenodo (DOI 10.5281/zenodo.23583135).

\bmhead{Competing interests}
The author declares no competing interests.

\bmhead{Author contributions}
H.H.: Conceptualization, Methodology, Software, Formal analysis, Investigation, Writing -- original draft, Writing -- review \& editing, Visualization, Funding acquisition.

\bibliography{refs}

\ifappendSI
\clearpage
\section*{Supplementary information}
\setcounter{figure}{0}
\setcounter{table}{0}
\renewcommand{\thefigure}{\arabic{figure}}
\renewcommand{\thetable}{\arabic{table}}
\renewcommand{\figurename}{Supplementary Fig.}
\renewcommand{\tablename}{Supplementary Table}

\begin{figure}[htbp]
\centering
\includegraphics[width=\textwidth]{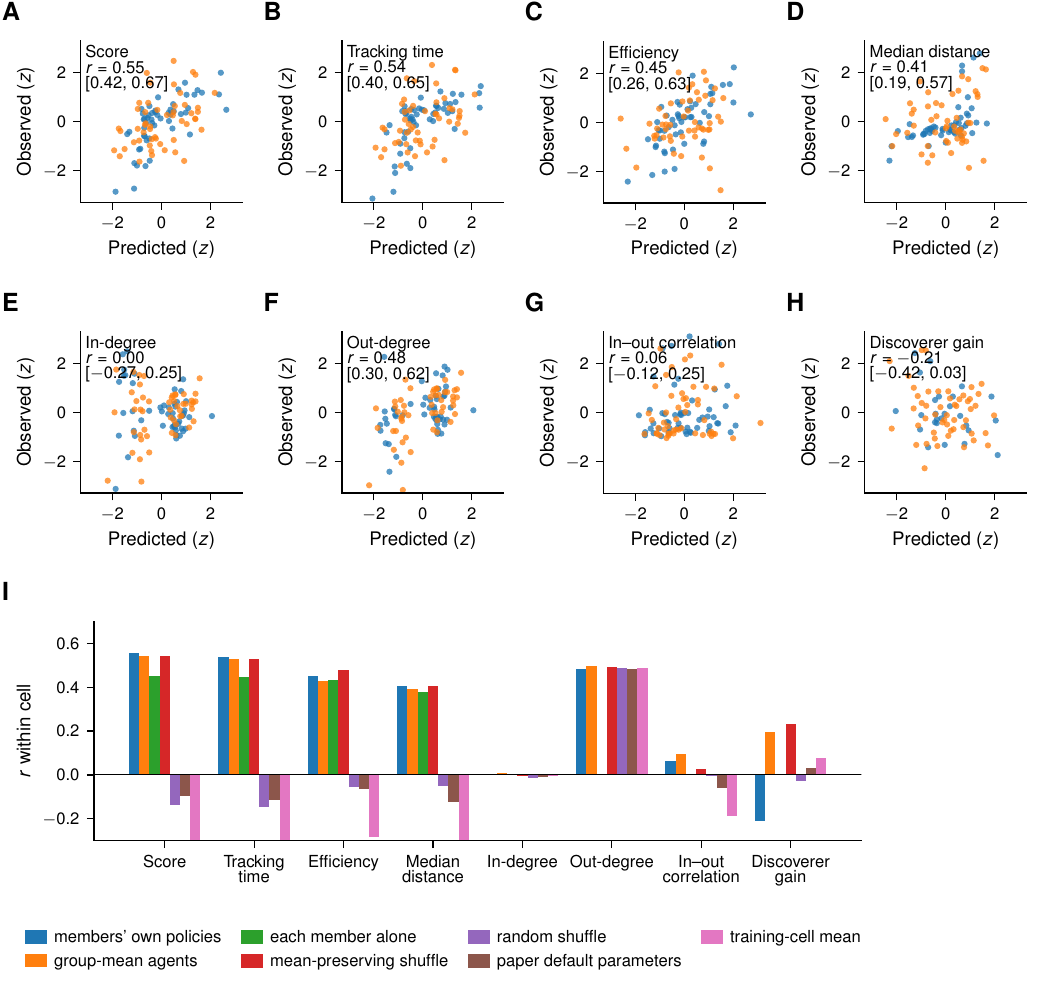}
\caption{\textbf{Chirkov et al.: leave-one-group-out prediction of all collective metrics by the individually fitted ABM policies and comparison with the controls.}
(A--H) Observed (vertical) against predicted (horizontal; mean of 100 seeds with the members' individual policies) score, tracking time, tracking efficiency (points per second of tracking), median distance from the resource centre, in-degree, out-degree, between-member in--out correlation and discoverer attention gain.
All standardised within condition $\times$ speed. Blue slow, orange fast. Each panel shows $r$ with its 95\,\% group-bootstrap confidence interval; $n = 103$ groups (discoverer attention gain: 77 groups with discovery events).
(I) Within-cell $r$ of each simulation variant: members' own policies; group-mean agents (all members replaced by the group's mean parameters); each member alone; mean-preserving re-composition; random re-composition within the same cell; the original study's default parameters; the within-cell mean parameters of the training fold.
The out-degree $r \approx 0.48$ shared by all variants reflects the number of analysed members (groups of four vs five) and carries no policy information.}
\label{fig:s1}
\end{figure}

\begin{figure}[htbp]
\centering
\includegraphics[width=\textwidth]{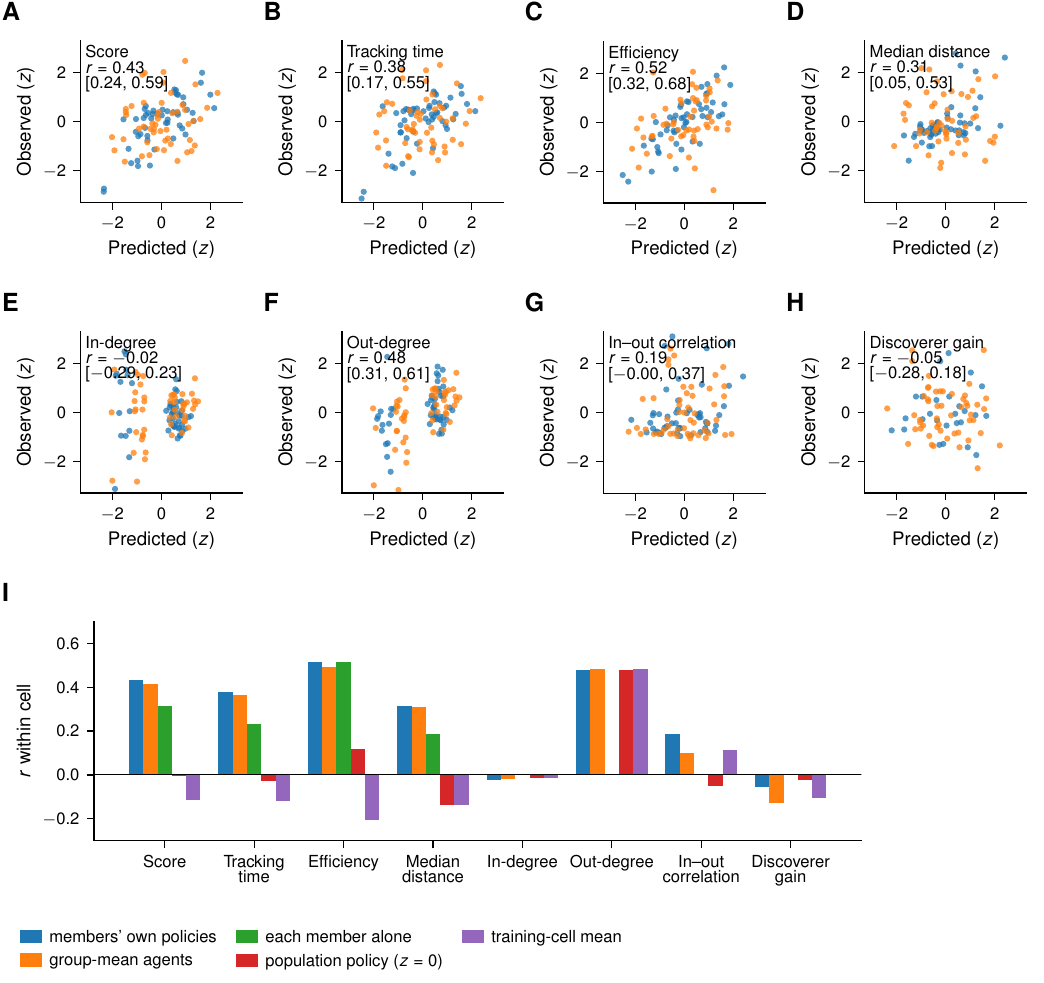}
\caption{\textbf{Chirkov et al.: the same panels for the neural policy conditioned on the participant embedding $\bm{e}$ (vectorised environment).}
(A--H) Predictions of the neural policy in the same layout as Supplementary Fig.~1.
(I) Within-cell $r$ of each variant: members' own $\bm{e}$; group-mean $\bm{e}$; each member alone; the population policy ($\bm{e} = 0$); the within-cell mean $\bm{e}$ of the training fold.
Score $r = 0.43$ \ci{0.24}{0.59}, tracking time 0.38, efficiency 0.52; difference from the group-mean-agent control for score $\dr = +0.02$ \ci{-0.07}{+0.12}.}
\label{fig:s2}
\end{figure}

\begin{figure}[htbp]
\centering
\includegraphics[width=0.75\textwidth]{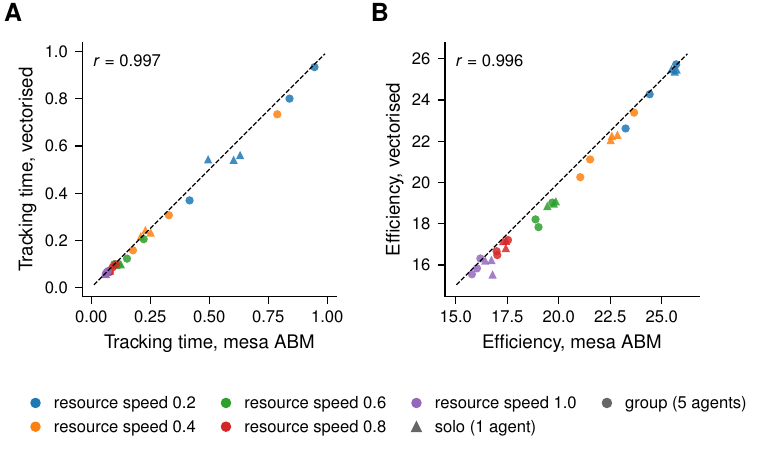}
\caption{\textbf{Agreement between the vectorised environment and the original mesa ABM.}
Homogeneous agents with the original study's default parameters were run on the grid of Fig~5B of Chirkov et al.\ (social-information sensitivity $k \in \{0.1, 1, 10\}$, resource speed 0.2--1.0; solitary agents as triangles, groups of five as circles; colour = resource speed).
(A) Tracking time and (B) tracking efficiency (points per tracking step), mesa ABM (horizontal; 100 runs per cell) against the numpy re-implementation used for the neural policy and the search over social rules (vertical; 200 runs per cell). Dashed: identity. $r = 0.997$ and 0.996.}
\label{fig:s3}
\end{figure}

\begin{figure}[htbp]
\centering
\includegraphics[width=\textwidth]{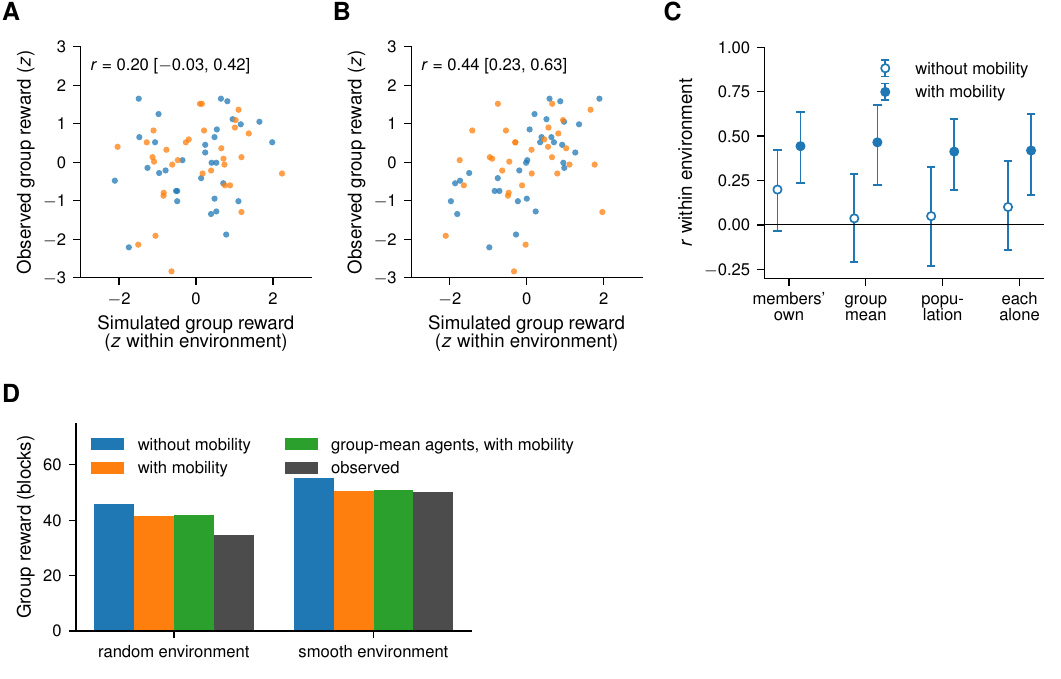}
\caption{\textbf{Wu et al.: the original grid ABM without and with a per-agent mobility term.}
(A) Observed group reward (mean of the four group rounds per session $\times$ environment, 64 cells, standardised within environment) against the zero-shot simulation with the members' fitted choice weights and the original one-cell-per-step movement ($r = 0.20$).
(B) The same with each agent moving on a step with probability $p_i = $ (mean solo-round path length) / (maximum over participants), calibrated from the solo rounds only ($r = 0.44$).
Blue random, orange smooth environment.
(C) Within-environment $r$ (session-cluster bootstrap 95\,\% confidence interval) of the four variants (members' own weights, group-mean agents, population weights, each member alone). Open: without the mobility term; filled: with it.
(D) Absolute level of group reward per environment: without mobility, with mobility, group-mean agents (with mobility), observed.}
\label{fig:s4}
\end{figure}

\begin{figure}[htbp]
\centering
\includegraphics[width=0.75\textwidth]{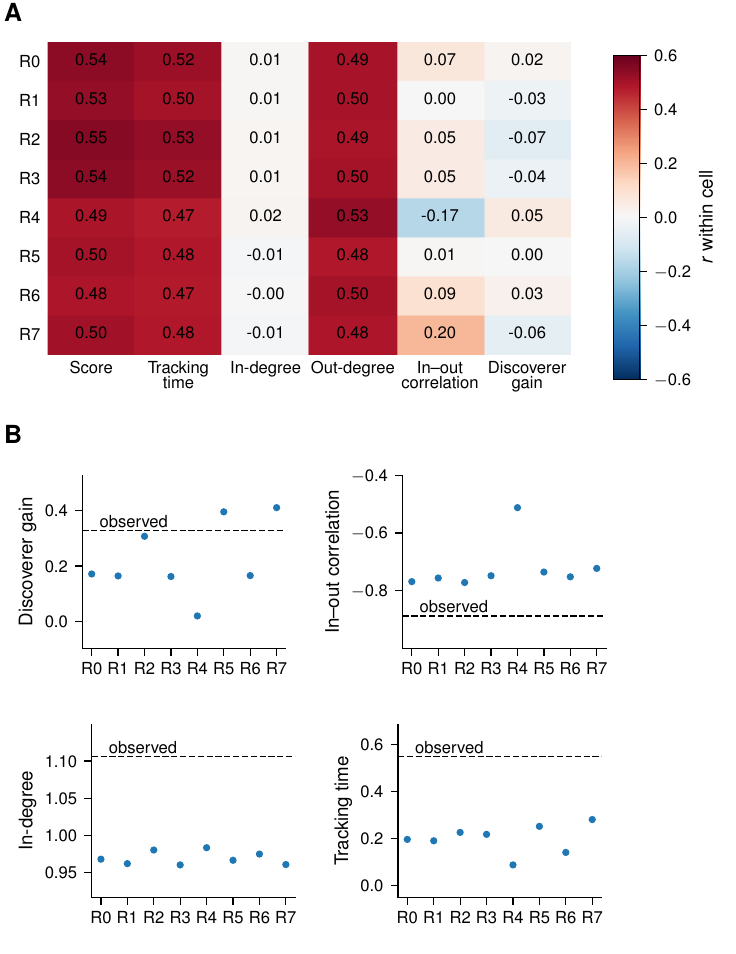}
\caption{\textbf{Chirkov et al.: search over social rules in the vectorised environment.}
The 103 analysed groups were each simulated 50 times with their members' fitted ABM parameters under the baseline rule (R0) and seven rules that change a single element of the social part of the decision rule: R1 turn towards the nearest visible other instead of the mean bearing of visible others; R2 judge the quality of social information from the best-informed visible other; R3 wide scanning turns ($p = 0.2$) while searching; R4 align with the others' mean heading instead of turning towards them; R5 use only others on the resource as a source of social information; R6 scale rotation while tracking by 0.8; R7 = R5 + R3.
(A) Within-cell $r$ between the seed-mean simulation and the observed group metrics (colour and number; colour bar at right).
(B) Levels (mean over groups) of discoverer attention gain, in--out correlation, in-degree and tracking time for R0--R7 (points) and observed (dashed).
Rules that direct attention to others on the resource (R2, R5, R7) reproduce the level of attention to discoverers (0.31--0.41 vs 0.33 observed), but no rule reproduces the between-group variation of the network metrics.}
\label{fig:s5}
\end{figure}

\clearpage
\begin{table}[htbp]
\centering
\caption{\textbf{Meaning of the individual parameters and their trait classes.}
RE = participant-level random effect of the original study's computational model (teacher model). Behavioural measures are summary statistics. Trait classes correspond to the colours of Figs~2--3.}
\label{tab:params}
\footnotesize
\setlength{\tabcolsep}{3pt}
\begin{tabular}{p{1.8cm}p{3.1cm}p{5.3cm}p{2.0cm}}
\toprule
Dataset & Parameter & Meaning & Trait class \\
\midrule
Chirkov (RE) & Intercept & Baseline amount of rotation while tracking (motor activity) & Private processing \\
 & Detector & Sensitivity to the resource detector (change in distance to the resource); strength of heading correction when about to lose the resource & Private processing \\
 & SocPosInFOV & Weight on turning towards the mean position of others in view & Social attraction \\
 & SocMovDir & Weight on aligning with the others' mean movement direction & Social attraction \\
 & SocPos$\times$SocQual, SocMov$\times$SocQual & Increment of the above weights when visible others are closer to the resource than oneself (payoff selectivity) & Selective social learning \\
Chirkov (ABM policy) & $I$, $S$ & Baseline rotation while tracking and sensitivity to the detector & Private processing \\
 & $k$, $x_0$ & Slope and threshold of the sigmoid that decides whether to follow others according to the quality of social information & Selective social learning \\
 & $r_{\max}$ & Upper bound of random rotation while searching & Private processing \\
 & $\pi$ (persistence) & Probability of not rotating in a second (intermittency of movement) & Private processing \\
Chirkov (roles) & in-degree / out-degree & Number of others who have one in their 90$^\circ$ field of view / number of others one has in view (mean per second) & Being attended to / attending to others \\
Wu (choice weights) & Locality & Tendency to choose nearby blocks (locality of search) & Private processing \\
 & BlockVisibility & Tendency to choose blocks in view & Private processing \\
 & RewardPrediction & Weight on the reward predicted from one's own reward experience (asocial learning) & Private processing \\
 & SuccessfulProximity & Weight on choosing near a player just seen to be rewarded & Selective social learning \\
 & UnsuccessfulProximity & Weight on choosing near a player seen not to be rewarded & Selective social learning \\
 & Adapt* & Change of each weight with the time since one's own last reward & Selective social learning \\
Wu (behaviour) & Path length, speed, cells traversed & Mobility in the solo rounds & Private processing (input to $z$) \\
 & out-degree, nearest-neighbour distance & Time with others in view, distance to others & Social attraction \\
 & in-degree & Time in the others' field of view & Being attended to \\
 & Number of rewards & Reward blocks per round & Performance \\
Hawkins (behaviour) & Stop latency, fraction stopped inside the region & Time from entering the scoring region to stopping; fraction of time stopped inside the region & Private processing \\
 & Approach, nearest-neighbour and centroid distance, facing others & Fast approach to one other; distance to others; time facing others (45$^\circ$ cone) & Social attraction \\
 & Facing stopped others, selectivity index & Time facing stopped others; degree to which stopped others are chosen over moving others & Selective social learning \\
 & Faced by others & Time being faced by others & Being attended to \\
Hawkins (teacher) & $b_{\mathrm{cen}}$, $b_{\mathrm{nn}}$ & Attraction to the others' centroid and to the nearest other & Social attraction \\
 & $b_{\mathrm{stop}}$ & Attraction to the nearest stopped other (inference of payoff) & Selective social learning \\
\bottomrule
\end{tabular}
\end{table}

\begin{table}[htbp]
\centering
\caption{\textbf{Split-half reliability of the individual parameters (all parameters).}
Chirkov: random effects of the tracking model (first vs second half, empirical-Bayes teacher, $n = 611$) and ABM policy parameters ($n = 621$). Wu: rounds 1--2 vs 3--4 of each block ($n = 128$). Hawkins: first vs second half, within experiment $\times$ group size $\times$ field (Experiment 1 $n = 557$--682, Experiment 3 $n = 338$--450).}
\label{tab:s1}
\scriptsize
\setlength{\tabcolsep}{3pt}
\begin{tabular}{p{1.5cm}p{3.2cm}>{\raggedleft\arraybackslash}p{0.8cm}@{\hspace{5mm}}p{1.5cm}p{3.2cm}>{\raggedleft\arraybackslash}p{0.8cm}}
\toprule
Dataset & Parameter & $r$ & Dataset & Parameter & $r$ \\
\midrule
Chirkov (RE) & Intercept & 0.76 & Wu (choice weights) & Locality & 0.83 \\
 & Detector & 0.60 & & BlockVisibility & 0.76 \\
 & SocPosInFOV & 0.30 & & RewardPrediction & 0.43 \\
 & SocMovDir & 0.17 & & SuccessfulProximity & 0.45 \\
 & SocPosInFOV:SocQual & 0.12 & & UnsuccessfulProximity & 0.27 \\
 & SocMovDir:SocQual & 0.09 & & AdaptLocality & 0.39 \\
Chirkov (ABM policy) & $I$ & 0.58 & & AdaptSuccessful & 0.20 \\
 & $S$ & 0.62 & & AdaptUnsuccessful & 0.16 \\
 & $k$ & 0.70 & Wu (behaviour) & Solo path length & 0.87 \\
 & $x_0$ & 0.94 & & Solo rewards & 0.54 \\
 & $r_{\max}$ & 0.59 & & Group rewards & 0.56 \\
 & $\sigma$ & 0.26 & & out-degree & 0.59 \\
 & persistence $\pi$ & 0.84 & & in-degree & 0.24 \\
Hawkins Exp.\ 1 & Stopped inside region & 0.64 & & NN distance & 0.37 \\
 & Stop latency & 0.69 & & Splashes observed & 0.46 \\
 & Facing others & 0.34 & Hawkins Exp.\ 3 & Facing others & 0.45 \\
 & Facing stopped others & 0.55 & & Facing stopped others & 0.59 \\
 & Selectivity index & 0.25 & & Selectivity index & 0.42 \\
 & Faced by others & 0.22 & & Faced by others & 0.29 \\
 & Approach & 0.61 & & Approach & 0.63 \\
 & NN distance & 0.59 & & NN distance & 0.61 \\
 & Centroid distance & 0.57 & & Centroid distance & 0.63 \\
 & Score & 0.41 & & Score & 0.64 \\
 & Teacher $b_{\mathrm{cen}}$ & 0.43 & & Teacher $b_{\mathrm{cen}}$ & 0.27 \\
 & Teacher $b_{\mathrm{stop}}$ & 0.16 & & Teacher $b_{\mathrm{stop}}$ & 0.23 \\
\bottomrule
\end{tabular}
\end{table}

\begin{table}[htbp]
\centering
\caption{\textbf{Wu et al.: prediction of group reward by the grid ABM without and with the mobility term.}
Top: within-environment $r$ of each variant (64 session $\times$ environment cells; session-bootstrap 95\,\% confidence interval) and $r$ by environment.
Middle: $\dr$ between variants.
Bottom: absolute level of group reward per environment (simulation vs observed).
Own = the participants' own choice weights, group mean = group-mean choice weights and mobility, population = population choice weights, alone = each member simulated alone; ``+ mobility'' = with the own (group-mean) mobility term.}
\label{tab:s3}
\scriptsize
\setlength{\tabcolsep}{3pt}
\begin{tabular}{p{3.6cm}*{4}{>{\raggedleft\arraybackslash}p{2.0cm}}}
\toprule
Variant & $r$ & 95\,\% CI & $r$ (random) & $r$ (smooth) \\
\midrule
Own & 0.20 & $[-0.03, 0.42]$ & 0.16 & 0.24 \\
Group mean & 0.04 & $[-0.21, 0.29]$ & $-0.03$ & 0.11 \\
Population & 0.05 & $[-0.23, 0.33]$ & 0.06 & 0.04 \\
Alone & 0.10 & $[-0.14, 0.36]$ & 0.13 & 0.08 \\
Own + mobility & 0.44 & $[0.24, 0.64]$ & 0.67 & 0.22 \\
Group mean + mobility & 0.47 & $[0.23, 0.67]$ & 0.64 & 0.29 \\
Population + mobility & 0.41 & $[0.20, 0.60]$ & 0.58 & 0.25 \\
Alone + mobility & 0.42 & $[0.17, 0.62]$ & 0.51 & 0.33 \\
\midrule
Contrast & $\dr$ & 95\,\% CI & $p$ & \\
\midrule
Own + mobility $-$ own & $+0.24$ & $[+0.04, +0.46]$ & 0.021 & \\
Group mean + mobility $-$ group mean & $+0.43$ & $[+0.17, +0.70]$ & 0.003 & \\
Population + mobility $-$ population & $+0.36$ & $[+0.06, +0.67]$ & 0.023 & \\
Alone + mobility $-$ alone & $+0.32$ & $[+0.11, +0.50]$ & 0.006 & \\
Own + mobility $-$ group mean + mobility & $-0.02$ & $[-0.16, +0.12]$ & 0.80 & \\
Own + mobility $-$ population + mobility & $+0.03$ & $[-0.14, +0.22]$ & 0.71 & \\
Own + mobility $-$ alone + mobility & $+0.02$ & $[-0.11, +0.15]$ & 0.73 & \\
\midrule
Environment & Own & Own + mobility & Group mean + mobility & Observed \\
\midrule
random & 45.7 & 41.5 & 41.9 & 34.8 \\
smooth & 55.4 & 50.4 & 51.0 & 50.1 \\
\bottomrule
\end{tabular}
\end{table}

\begin{table}[htbp]
\centering
\caption{\textbf{Composition test with selectivity measured inside the group.}
Out-of-sample $\dr$ (cluster-bootstrap 95\,\% confidence interval and $p$) and in-sample $\Delta R^2$ (permutation $p$ of the SD block) when the within-group SD of selectivity measured inside the group is added to the model of member means (baseline).
Wu: SD of the weights fitted on the first two rounds of each block → group reward of the last two rounds (128 group rounds); baseline $r = 0.14$ \ci{-0.09}{0.36}.
Hawkins: SD of the first-half measures → second-half group score (169 games with at least three players); baseline $r = 0.45$ \ci{0.33}{0.57}.}
\label{tab:s2}
\footnotesize
\setlength{\tabcolsep}{3pt}
\begin{tabular}{llrrrr}
\toprule
Dataset & SD added & $\dr$ & 95\,\% CI & $p$ & $\Delta R^2$ ($p$) \\
\midrule
Wu & SuccessfulProximity & $+0.007$ & $[-0.065, +0.047]$ & 0.41 & 0.029 (0.041) \\
Wu & Both social weights & $-0.006$ & $[-0.079, +0.072]$ & 0.55 & 0.034 (0.10) \\
Wu & All 5 weights & $-0.051$ & $[-0.116, +0.026]$ & 0.89 & 0.053 (0.18) \\
Hawkins & Facing stopped others & $-0.004$ & $[-0.023, +0.014]$ & 0.63 & 0.004 (0.34) \\
Hawkins & Selectivity index & $-0.005$ & $[-0.020, +0.011]$ & 0.72 & 0.008 (0.17) \\
Hawkins & Teacher $b_{\mathrm{stop}}$ & $-0.006$ & $[-0.019, +0.007]$ & 0.81 & 0.000 (0.81) \\
Hawkins & All 4 measures & $-0.020$ & $[-0.052, +0.009]$ & 0.91 & 0.015 (0.50) \\
\bottomrule
\end{tabular}
\end{table}

\begin{table}[htbp]
\centering
\caption{\textbf{Pre-specified hypotheses, decision rules and verdicts, and deviations from the specification.}
Transcribed from the analysis specification fixed before the analysis and from the dated decision log kept during the analysis.}
\label{tab:s4}
\footnotesize
\setlength{\tabcolsep}{3pt}
\begin{tabular}{p{1.5cm}p{5.0cm}p{5.8cm}}
\toprule
Hypothesis & Decision rule (pre-specified) & Result \\
\midrule
(i) Search → tracking & The $z$ of a sequence encoder of the search phase predicts the tracking-phase random effects with $\dr \ge 0.10$, $p < 0.01$ against the condition $\times$ speed mean, and outperforms a linear ridge & \textbf{Failed}. Mean $r = 0.09$ for the CNN vs 0.17 for ridge. The condition-mean baseline gives $r \approx 0$ by the definition of random effects and is uninformative ($\dr = 0.47$ is trivial). Unchanged with a GRU, $M = 2, 8$, six learning rates $\times$ 500/10k/50k epochs \\
(ii) Cross-individual transfer & Distance between individuals in $z$ space explains transfer error (slope $p < 0.001$) and outperforms the summary-statistic distance & Held (linear $z$: slope 0.124 for the $z$ distance vs 0.087 for the summary-statistic distance; log-likelihood $-23{,}883$ vs $-24{,}505$) \\
(iii) Early movement → roles & The first-3-min $z$ predicts later roles with $\dr \ge 0.05$, $p < 0.01$ against the initial-score baseline & In-degree \textbf{held} ($\dr = 0.114$, group-cluster bootstrap $p = 0.007$); out-degree \textbf{failed} (0.079, $p = 0.14$). The reading (testing $\dr$ directly by cluster bootstrap) was decided on 2026-10-20 \\
(iv) Payoff information and roles & Prediction is best in the FP condition & Not supported (in-degree $r$: NP 0.18 / VP 0.27 / FP 0.18) \\
Payoff sharing & Exploratory (no verdict) & The fraction shared is exhausted by tracking time (Supplementary Note 1) \\
(v) Individual → group performance & Agents with individual parameters predict score and efficiency with $\dr \ge 0.10$, $p < 0.01$ against the default-parameter and mean-parameter controls & Held (score $\dr = 0.65$ and 0.93). Both controls, however, are constant within cell ($r \approx 0$) and trivial \\
(vi) Composition & Reproduction is more accurate than random re-composition of the groups (permutation $p < 0.01$) & Random re-composition: held ($p < 0.001$; trivial because re-composition changes the mean). Mean-preserving re-composition (added): failed (score $p = 0.012$, efficiency 0.19). Substantive test = the group-mean-agent control (decided 2026-10-20): $\dr = +0.01$ \ci{-0.03}{+0.06} \\
\midrule
Deviations & \multicolumn{2}{p{11.0cm}}{Teacher estimated by empirical Bayes on the identical model rather than brms (brms $> 1$ day per model; $r$ 0.91--0.998 confirmed on a 20\,\% subsample). Angle normalisers are geometric constants. The 53 participants with less than 10 s of search phase were excluded from the transfer analysis. ABM policies fitted directly per participant rather than by a hypernetwork. $\sigma$ fixed at the original study's 10$^\circ$, $\pi$ the group mean. Seeds 200 → 100, re-compositions 100 → 50 $\times$ 10 seeds. Groups of four simulated with four agents. Control 3 is the simulation of each member alone. The evaluation of the exploratory analysis of payoff sharing (Supplementary Note 1) changed from leave-one-participant-out cross-validation + Pearson (an artefact giving $r \approx -1$ for constant predictions) to fixed 10 folds + Spearman.} \\
\bottomrule
\end{tabular}
\end{table}

\clearpage
\subsection*{Supplementary Note 1. Exploratory analysis: voluntary payoff sharing (Chirkov et al., VP condition)}
In the VP condition (165 participants), participants could choose at any moment to show their own payoff information to the others at a cost of 1 point per second.
As an exploratory analysis without pre-specified decision criteria, we predicted the fraction of time sharing (overall, while tracking, while searching), the probability of sharing within 10 s of entering the resource (median 13 entries per participant), and the probability of sharing within 10 s of discovering the resource oneself ($n = 88$) from the linear search-phase $z$, the first-3-min $z$, the tracking-phase random effects (6 coefficients) and trivial controls (speed, score, tracking time), by ridge regression (logit-transformed fractions) in fixed 10 folds of whole groups.
The pre-specified leave-one-participant-out cross-validation with Pearson $r$ produced an artefact of $r \approx -1$ for weak predictors (an inverse correlation with the mean of the held-out participants), so it was changed to fixed 10 folds and Spearman $\rho$ (participant-bootstrap 95\,\% confidence interval) (Supplementary Table 6).

\begin{table}[htbp]
\centering
\caption{\textbf{Out-of-fold prediction of voluntary payoff sharing (Spearman $\rho$, participant-bootstrap 95\,\% confidence interval).}}
\label{tab:s5}
\scriptsize
\setlength{\tabcolsep}{2pt}
\begin{tabular}{p{2.2cm}*{5}{>{\raggedleft\arraybackslash}p{1.98cm}}}
\toprule
Target & Score only & Tracking time only & Search-phase $z$ & First-3-min $z$ & Tracking RE \\
\midrule
Fraction shared (overall; $n = 165$) & 0.23 $[0.08, 0.37]$ & 0.29 $[0.14, 0.42]$ & 0.23 $[0.08, 0.38]$ & 0.16 $[-0.00, 0.31]$ & 0.32 $[0.18, 0.46]$ \\
Fraction shared (while tracking) & $-0.34$ $[-0.47, -0.20]$ & $-0.30$ $[-0.44, -0.16]$ & $-0.11$ $[-0.26, 0.05]$ & $-0.10$ $[-0.25, 0.06]$ & 0.10 $[-0.05, 0.24]$ \\
Fraction shared (while searching; $n = 162$) & $-0.01$ $[-0.15, 0.15]$ & $-0.14$ $[-0.30, 0.02]$ & 0.09 $[-0.08, 0.24]$ & $-0.17$ $[-0.32, -0.00]$ & 0.02 $[-0.12, 0.17]$ \\
Sharing after entering the resource ($n = 163$) & 0.03 $[-0.12, 0.19]$ & 0.07 $[-0.09, 0.22]$ & 0.21 $[0.06, 0.37]$ & 0.14 $[-0.02, 0.29]$ & 0.16 $[0.01, 0.29]$ \\
Sharing after one's own discovery ($n = 88$) & $-0.13$ $[-0.34, 0.10]$ & 0.04 $[-0.16, 0.25]$ & 0.10 $[-0.12, 0.31]$ & 0.21 $[0.00, 0.42]$ & $-0.18$ $[-0.37, 0.04]$ \\
\bottomrule
\end{tabular}
\end{table}

Reading: the overall fraction shared is predicted no better than by the trivial controls (tracking time $\rho = 0.29$, score 0.23), and because sharing is possible only while scoring, it is exhausted by ``people who track longer share more'' (the raw Spearman correlations of the fraction shared with score and tracking time are 0.34 / 0.35).
The movement-style $z$ exceeds the controls only for sharing immediately after entering the resource (search-phase $z$ $\rho = 0.21$ \ci{0.06}{0.37} vs score 0.03 and tracking time 0.07).
Sharing while tracking and while searching is predicted by none of the predictors (a negative $\rho$ means ``no signal'').
The effects are small and exploratory, and the main text mentions them in a single sentence.

\fi

\end{document}